\documentclass[10pt,twocolumn,letterpaper]{article}

\usepackage{cvpr}              

\usepackage[dvipsnames]{xcolor}

\definecolor{cvprblue}{rgb}{0.21,0.49,0.74}

\usepackage{graphicx}
\usepackage{amsmath}
\usepackage{amssymb}
\usepackage{booktabs}

\usepackage{color}
\usepackage{amssymb}
\usepackage{lipsum}
\usepackage{algorithm}
\usepackage{algpseudocode}
\usepackage{xcolor}
\usepackage{tabularx}
\usepackage{multirow}
\usepackage{enumitem}
\usepackage{bbm}
\usepackage{wrapfig}
\usepackage{setspace}
\usepackage{footnote}

\usepackage{subcaption}
\usepackage{colortbl} 
\usepackage{pifont}  
\usepackage{cite}  
\usepackage{mathtools}  
\usepackage[font=small]{caption}  
\usepackage{blindtext}  
\usepackage{stmaryrd} 
\usepackage[accsupp]{axessibility}  
\usepackage{mathtools}
\usepackage{bigdelim}

\usepackage{subfiles}
\usepackage{afterpage}

\usepackage{algorithm}
\usepackage{algpseudocode}
\usepackage{adjustbox}

\usepackage[pagebackref,breaklinks,colorlinks,citecolor=cvprblue]{hyperref}

\usepackage{afterpage}
\usepackage{ulem} 
\usepackage{xcolor,pifont}

\definecolor{amethyst}{rgb}{0.6, 0.4, 0.8}

\definecolor{grey}{rgb}{0.9, 0.9, 0.9}
\newcommand{\ccol}{\cellcolor{grey}}
\newcommand*\colourcheck[1]{%
  \expandafter\newcommand\csname #1check\endcsname{\textcolor{#1}{\ding{51}}}%
}

\colourcheck{blue}
\colourcheck{green}
\colourcheck{red}

\def\ie{\emph{i.e.}}
\def\eg{\emph{e.g.}}

\definecolor{bright_red}{rgb}{0.97, 0.9, 0.9}
\definecolor{blk}{rgb}{0, 0, 0}
\definecolor{grn}{rgb}{0, 0.6, 0}
\definecolor{mgt}{rgb}{0.8, 0.1, 0.8}
\definecolor{darkblue}{rgb}{0.2, 0.2, 0.8}
\definecolor{lblue}{rgb}{0.2, 0.2, 1.0}
\definecolor{orange}{rgb}{1.0, 0.5, 0.2}
\definecolor{goldenrod}{rgb}{0.85, 0.65, 0.13}

\newcommand*{\bigs}[1]{\scalebox{1.05}{\ensuremath#1}}

\makeatletter
\newcommand{\multiline}[1]{%
  \begin{tabularx}{\dimexpr\linewidth-\ALG@thistlm}[t]{@{}X@{}}
    #1
  \end{tabularx}
}
\makeatother

\usepackage[capitalize]{cleveref}
\crefname{section}{Sec.}{Secs.}
\Crefname{section}{Section}{Sections}
\Crefname{table}{Table}{Tables}
\crefname{table}{Tab.}{Tabs.}

\def\ie{\emph{i.e.}}
\def\eg{\emph{e.g.}}

\definecolor{brown}{rgb}{0.85, 0.15, 0.15}
\definecolor{purp}{rgb}{0.95, 0.16, 0.65}
\definecolor{purpc}{rgb}{0.95, 0.36, 0.65}
\definecolor{orange}{rgb}{1.0, 0.5, 0.0}
\definecolor{blue}{rgb}{0.0, 0.5, 1.0}
\definecolor{green}{rgb}{0, 0.8, 0}
\definecolor{lgreen}{rgb}{0.6, 0.8, 0}
\definecolor{red}{rgb}{0.8, 0, 0}
\definecolor{darkblue}{rgb}{0.2, 0.2, 0.8}
\definecolor{brinkpink}{rgb}{0.98, 0.38, 0.5}
\definecolor{cadmiumred}{rgb}{0.89, 0.0, 0.13}
\definecolor{ceruleanblue}{rgb}{0.16, 0.32, 0.75}
\definecolor{dandelion}{rgb}{0.94, 0.88, 0.19}
\definecolor{bostonuniversityred}{rgb}{0.8, 0.0, 0.0}
\definecolor{brown(web)}{rgb}{0.65, 0.16, 0.16}
\definecolor{cornellred}{rgb}{0.7, 0.11, 0.11}

\newcolumntype{C}[1]{>{\centering\let\newline\\\arraybackslash\hspace{0pt}}p{#1}}

\definecolor{grey}{rgb}{0.9, 0.9, 0.9}

\DeclareMathOperator*{\argmin}{argmin} 
\DeclareMathOperator*{\argmax}{argmax} 

\newcommand{\hyperfootnote}[1][]{\def\ArgI\hyperfootnoteRelay}
\newcommand\hyperfootnoteRelay[2][]{\href{#1#2}{\ArgI}\footnote{\href{#1#2}{#2}}}

\def\paperID{5380} 
\def\confName{CVPR}
\def\confYear{2025}

\title{GENIUS: A Generative Framework for Universal Multimodal Search}

\author{
Sungyeon Kim$^{1,2}$ \,
Xinliang Zhu$^1$ \,
Xiaofan Lin$^1$ \,
Muhammet Bastan$^1$ \,
Douglas Gray$^1$ \,
Suha Kwak$^2$ \\[0.3em]
{$^1$ Amazon \ \ \ \ \ \ \ \ \ \ \ \ \ \ \ \ \ \ \ \ $^2$ POSTECH} \\
{\tt\small \{sungyeon.kim,suha.kwak\}@postech.ac.kr} \quad
{\tt\small \{xlzhu, xiaofanl, mbastan, douggray\}@amazon.com}
\vspace{-1mm}
}

\begin{document}

\maketitle

\setlength{\abovedisplayskip}{6.5pt} 
\setlength{\belowdisplayskip}{6.5pt}

\begin{abstract}

Generative retrieval is an emerging approach in information retrieval that generates identifiers (IDs) of target data based on a query, providing an efficient alternative to traditional embedding-based retrieval methods. However, existing models are task-specific and fall short of embedding-based retrieval in performance. This paper proposes GENIUS, a universal generative retrieval framework supporting diverse tasks across multiple modalities and domains. At its core, GENIUS introduces modality-decoupled semantic quantization, transforming multimodal data into discrete IDs encoding both modality and semantics. Moreover, to enhance generalization, we propose a query augmentation that interpolates between a query and its target, allowing GENIUS to adapt to varied query forms. Evaluated on the M-BEIR benchmark, it surpasses prior generative methods by a clear margin. Unlike embedding-based retrieval, GENIUS consistently maintains high retrieval speed across database size, with competitive performance across multiple benchmarks. With additional re-ranking, GENIUS often achieves results close to those of embedding-based methods while preserving efficiency.
\end{abstract}

\section{Introduction}
\label{sec:intro}

\begin{figure}[!t]
\centering
\includegraphics[width = 0.93\linewidth]{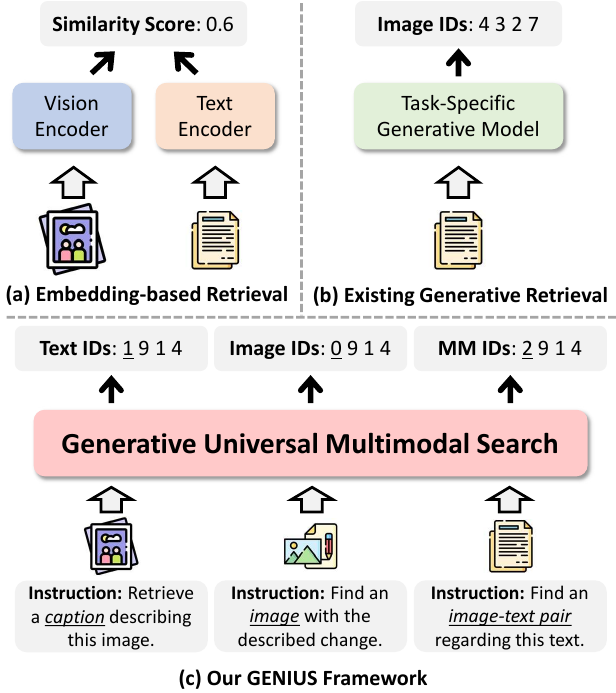}
\vspace{-2mm}
\caption{Illustrations of three Information Retrieval paradigms. (a) Embedding-based retrieval, where queries and candidates are embedded, and similarity is measured. (b) Existing generative retrieval generates task-specific identifiers. (c) The GENIUS framework generates identifiers across modalities based on queries and instructions, where the first-level code indicates modality.} 
\label{fig:thumbnail}
\vspace{-4mm}
\end{figure}

Information Retrieval (IR) is a fundamental task of finding relevant information from a large database~\cite{manning2009introduction, singhal2001modern}. 
With the rapid growth of data, efficient and accurate IR is more essential than ever.
Conventional IR approaches commonly follow the embed-and-retrieve paradigm, known as embedding-based retrieval~(Fig.~\ref{fig:thumbnail}(a)). They embed the query and the database into a high-dimensional embedding space, which is learned by metric learning~\cite{songCVPR16,Sohn_nips2016,Yu_2019_ICCV,wang2019multi,movshovitz2017no,kim2020proxy}, and then find the nearest neighbors of the query. As the database expands, however, a scalability issue arises due to the rapidly increasing cost of index building, maintenance, and nearest neighbor search, even with approximate nearest neighbor search like HNSW~\cite{malkov2018efficient} and Faiss~\cite{douze2024faiss}.

Recently, generative retrieval has emerged as a promising alternative. 
Inspired by Differentiable Search Index~\cite{tay2022transformer} and SPLADE~\cite{formal2021splade}, this approach generates identifiers (IDs) of target data directly from a query, bypassing the nearest neighbor search. 
However, existing methods in this line of research have limited capability due to their task-specific designs. Most of them are dedicated to text retrieval~\cite{tay2022transformer, wang2022nci}, and only a few recent works address images~\cite{zhang2023irgen} and cross-modal retrieval~\cite{li2024grace} (Fig.~\ref{fig:thumbnail}(b)). 
Hence, these methods fail to meet the diverse, multimodal demands of users in real-world applications.
Moreover, existing generative methods underperform in cross-modal retrieval compared to embedding-based retrieval methods~\cite{zhang2023irgen, li2024grace}.

\begin{figure*} [!t]
\centering
\includegraphics[width = 1.0\textwidth]{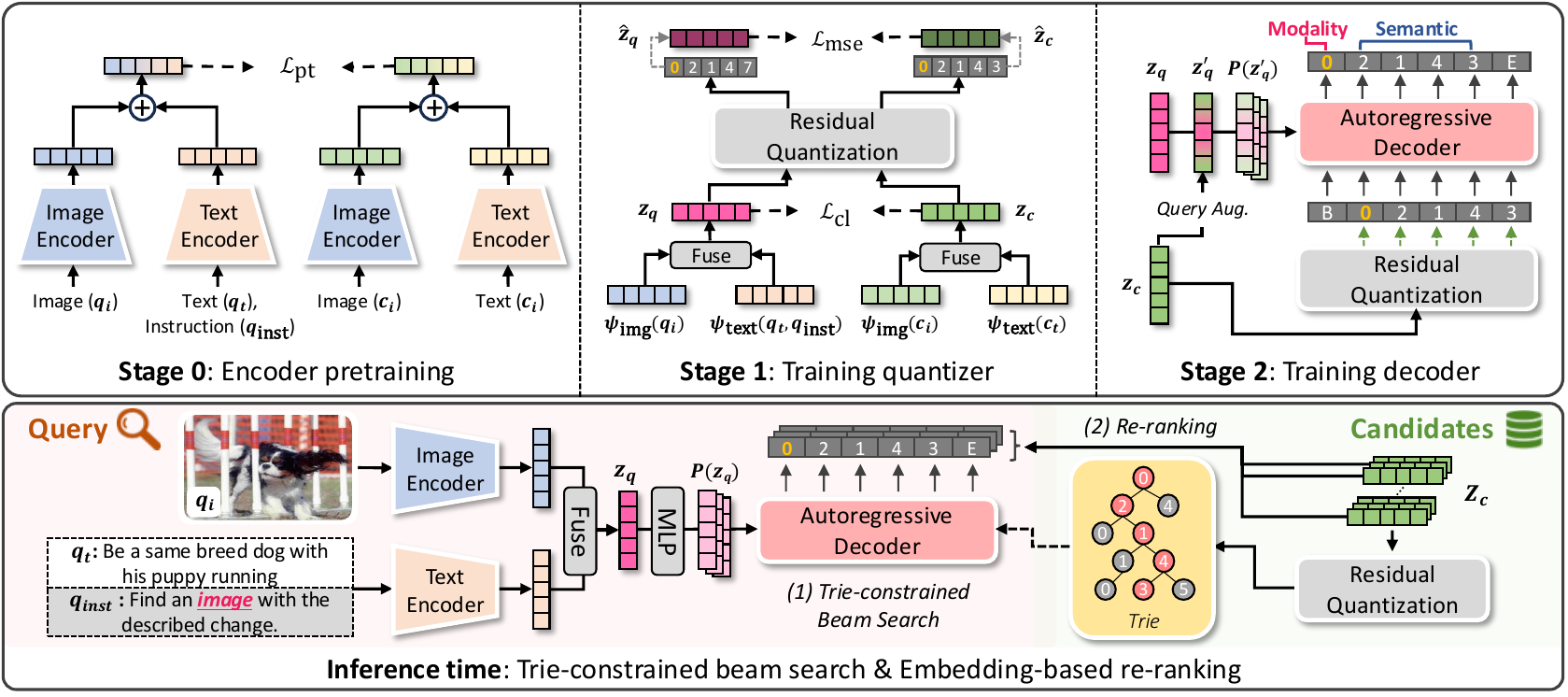}
\vspace*{-6mm}
\caption{ 
\textbf{Overview of the GENIUS framework.} GENIUS includes three components: image and text encoders, a modality-decoupled quantization module, and an autoregressive decoder. The framework follows three stages in training. First, the image-text encoders are pre-trained to enhance instruction comprehension and representation abilities. Next, residual quantization is trained to assign discrete IDs to candidate embeddings, where the first quantization level captures modality information and subsequent levels encode semantic details.  Finally, the decoder learns to generate modality-decoupled semantic IDs.
At inference, GENIUS generates candidate IDs from a query using Trie-constrained beam search, additionally followed by embedding-based re-ranking to further enhance retrieval accuracy.
}
\label{fig:overview}
\vspace*{-2mm}
\end{figure*}

In this paper, we propose \textbf{GENeratIve Universal multimodal Search} (GENIUS), the first generative retrieval framework that handles diverse retrieval tasks across multiple modalities. In GENIUS, each task is defined as finding data of a specified type, based on a multimodal query with an instruction that clarifies the user's intention.
Our framework uses the instructions to retrieve data of the appropriate format and domain among diverse data within the database. Unlike prior generative methods restricted to specific modalities or tasks, GENIUS generates IDs of relevant data across heterogeneous modalities, effectively addressing a wide range of retrieval scenarios. GENIUS consists of a multimodal encoder that processes the query and instruction, coupled with a decoder that generates target IDs based on this input, as illustrated in Fig.~\ref{fig:overview}.

A key contribution of GENIUS is \textbf{modality-decoupled semantic quantization} to assign a target ID to multimodal data. It transforms multimodal data into compact, layered representations capturing both semantic content and modality.
Fig.~\ref{fig:thumbnail}(c) illustrates this concept, with each target ID represented as a sequence of discrete codes comprising two components. The first code of the target ID indicates the data modality (\eg, 0 for images, 1 for text, and 2 for image-text pairs). This is achieved by training a quantization model with instructions that specify the modality of the target, allowing GENIUS to separate different modalities of the target. The subsequent codes capture the semantic content of the data while ensuring compatibility across modalities.
For example, when image and text have similar contents, their IDs should be similar, particularly in their leading codes (except the first one which is kept for modality encoding), regardless of their modality. This is achieved through contrastive learning combined with residual quantization, which clusters semantically related items, enabling a nuanced representation from coarse to fine granularity. 

Next, we train the decoder to generate target IDs from a given query. While these compact IDs are effective, they inherently contain less information than dense embeddings. As a result, the model may struggle to generalize to new or varied queries, especially with limited query-target pairs. To address this, we introduce \textbf{Query Augmentation} strategy.
This strategy generates augmented queries by linearly interpolating between the embeddings of a query and its corresponding target. Including these augmented queries in training enriches the data with diverse query examples that retain the same semantics. This augmentation allows the decoder to learn a more generalized mapping from queries to target IDs, making it robust to variations in query formulations at test time.

We train and evaluate GENIUS on a large-scale multimodal benchmark, M-BEIR~\cite{uniir}, which includes instructions for multimodal retrieval tasks. GENIUS outperforms the best generative retrieval method by 28.6 points in Recall@5 on the COCO dataset~\cite{Mscoco} for text-to-image retrieval.
Unlike prior generative models, GENIUS supports a broader range of tasks and significantly narrows the performance gap to embedding-based retrieval methods across multiple tasks.  It maintains a nearly constant retrieval speed across database sizes, and operates faster than previous generative methods. Moreover, by re-ranking predicted candidates based solely on their embeddings, GENIUS often achieves results close to those of embedding-based baselines in several tasks while preserving high efficiency. This combination of \emph{versatility}, \emph{performance}, and \emph{efficiency} marks a big step forward for generative multimodal retrieval.

\section{Related Work}
\label{sec:relate}


\subsection{Multimodal Information Retrieval}
Multimodal Information Retrieval (IR) has advanced significantly, particularly in cross-modal tasks like text-to-image retrieval. Traditional methods are divided into two main approaches: multi-encoder and single-encoder with cross-attention. Multi-encoder models~\cite{clip, girdhar2023imagebind, zhu2024bringing, align, zhai2023siglip, kim2023improving} efficiently map visual and textual features and other format features into a shared embedding space. Single-encoder models~\cite{li2022blip, wei2020universal, li2023blip2, li2020unicoder} provide more detailed modality interactions but incur a higher computational cost. 
Recent advances in IR have introduced composed image retrieval (CIR) tasks, which integrate image and text inputs based on user feedback~\cite{clip4cir, bai2023sentence, saito2023pic2word}. Fine-grained retrieval also requires models to handle complex multimodal queries, posing additional challenges~\cite{changpinyo2021telling, luo2023end}. Moreover, benchmarks like WebQA~\cite{chang2022webqa} and frameworks such as UniIR~\cite{uniir} extend IR capabilities to retrieve diverse data types, supporting unified retrieval across multiple datasets for broader generalization.
Most retrieval methods follow the embed-to-retrieve paradigm, while recent efforts~\cite{li2024grace, zhang2023irgen} have started to explore generative approaches for handling multi-modal tasks, which remain largely unexplored.

\subsection{Generative Retrieval}
Generative retrieval has recently emerged as an innovative paradigm, primarily targeting text-based document retrieval. Early works explored generating concise identifiers (IDs), such as entity names or passage titles, to represent documents effectively~\cite{de2020autoregressive, bevilacqua2022autoregressive}. These approaches have evolved into more generalized methods, such as NCI~\cite{wang2022neural} and DSI~\cite{tay2022transformer}, which use hierarchical clustering of document embeddings and pretrained language models to assign document identifiers effectively. Recent studies have further refined these concepts~\cite{du2024bottleneck, nguyen2023generative, tang2023semantic, mehta2022dsi++, rajput2023recommender}, with some proposing end-to-end methods to directly learn IDs~\cite{jin2023language, sun2024learning}.
While text retrieval benefits from the inherent discreteness of language, extending generative retrieval to multiple modalities introduces challenges in addressing modality gaps. GRACE~\cite{li2024grace} is one of the few studies that has explored cross-modal generative retrieval by introducing semantic IDs for images, while IRGen~\cite{zhang2023irgen} focuses solely on image-based retrieval and struggles with tasks beyond single-modality scenarios. These models are designed for a specific scenario and show significantly lower performance than embedding-based retrieval methods, highlighting their limitations in real-world applications. Our work addresses these limitations by introducing a \emph{universal framework} that dynamically generates IDs across text and images, supporting a broader range of retrieval tasks.


\section{Problem Formulation}
Universal multimodal search~\cite{uniir} aims to enable users to query and retrieve targets across diverse tasks based on user instruction $q_{\text{inst}}$. In this setup, we define
a query $\mathbf{q}$ as a combination of the query content and the instruction, represented as $({q_\text{con}}, q_{\text{inst}})$, where $q_\text{con}$ can take various forms, including an image $q_i$, text $q_t$, or an interleaved image-text pair $(q_i, q_t)$. The target candidate $\mathbf{c}$ can be represented as an image $c_i$, text $c_t$, or an interleaved image-text pair $(c_i, c_t)$.

We formalize universal generative multimodal search as the process of generating an ID $T_c$ for the relevant target $\mathbf{c}$, conditioned on the query $\mathbf{q}$:
\begin{equation}
\begin{gathered}
    T_{c} := (t^{c}_1, \ldots, t^{c}_M) \\
    \textrm{where} \,\,\, 
    t_k^c = \argmax_{t \in \mathcal{T}} \left[ \log p\left( t \mid \mathbf{q}, t^c_{<k}; \theta \right) \right],
\end{gathered}
\label{eq:gen_ret}
\end{equation}
where $\theta$ denotes the parameters of both the encoder and decoder, $t^c_{<k}$ is the previously generated tokens, and $p(\cdot)$ is the probability distribution over the next token given the context. That is, the model generates the ID $T_c$ by sequentially predicting tokens $t^c_k$ that maximize the conditional probability. This generative approach eliminates the need for similarity computations, indexing, and ranking across the entire target dataset, making retrieval efficient and scalable.

\section{Proposed Method}
\label{sec:method}

To address the universal generative retrieval problem, we propose GENeratIve Universal multimodal Search, dubbed GENIUS, which aims to generate target IDs across various modalities, guided by multimodal queries and instructions. 

As shown in Fig.~\ref{fig:overview}, GENIUS involves three distinct training stages. First, in Sec.~\ref{subsec:pretraining}, we describe multimodal encoder pretraining, which enables the encoder to effectively comprehend instructions and extract meaningful image-text features, aligning query intent with target semantics. Next, Sec.~\ref{subsec:quantizaiton} introduces the modality-decoupled quantization module, which quantizes multimodal embeddings into discrete IDs, explicitly encoding modality and semantic information. These discrete IDs then serve as target outputs for decoder training. Finally, Sec.~\ref{subsec:decoder} presents the autoregressive decoder training process, enabling the decoder to generate modality-decoupled semantic IDs directly from the query. In Sec.~\ref{subsec:inference}, we detail the inference pipeline of GENIUS. 

\subsection{Encoder Pretraining}
\label{subsec:pretraining}
To handle diverse retrieval tasks, a model should understand the relations between queries and targets by comprehending both query content and instructions. We achieve this through encoder pretraining, which enables the multimodal encoder to understand query semantics and instructive information.
For image and text encoders, we leverage CLIP~\cite{clip}. Specifically, we use the text encoder $\psi_{\text{text}}$ to process text-based query contents $q_t$ and instructions $q_{\text{inst}}$, while the image encoder $\psi_{\text{image}}$ is used for image inputs $q_i$.

To ensure strong alignment between queries and their corresponding positive targets, we employ contrastive learning. When both modalities are present in a query or target, we combine their features using simple element-wise addition~\cite{uniir, liu2022universal} to create a unified embedding: $\phi(\mathbf{q}) = \psi_{\text{image}}(q_i) + \psi_{\text{text}}(q_t, q_\text{inst})\in\mathbb{R}^{d}$ for query, and $\phi(\mathbf{c}) = \psi_{\text{image}}(c_i) + \psi_{\text{text}}(c_t)\in\mathbb{R}^{d}$ for targets, where $d$ is the embedding dimension. The contrastive loss between the query and target embeddings is defined as:
\begin{equation}
\mathcal{L}_{\text{pt}} = -\log \frac{\exp\left(\left< \phi(\mathbf{q}), \phi(c^+)\right> / \tau \right)}{\sum_{c' \in \mathcal{C}} \exp\left(\left< \phi(\mathbf{q}), \phi(c') \right> / \tau \right)},
\label{eq:pt_contra}
\end{equation}
where $\phi(c^+)$ is the embedding of a target corresponding to the query $\mathbf{\mathbf{q}}$, $\mathcal{C}$ is the set of all candidates, $\left<\cdot, \cdot\right>$ denotes cosine similarity, and $\tau$ is a temperature parameter. 
This training follows the CLIP-based learning framework of UniIR~\cite{uniir}. For implementation simplicity, we directly utilize its pre-trained weights.
After this phase, both the image and text encoders are frozen.

\begin{figure*} [!t]
\centering
\includegraphics[width = 1.0\textwidth]{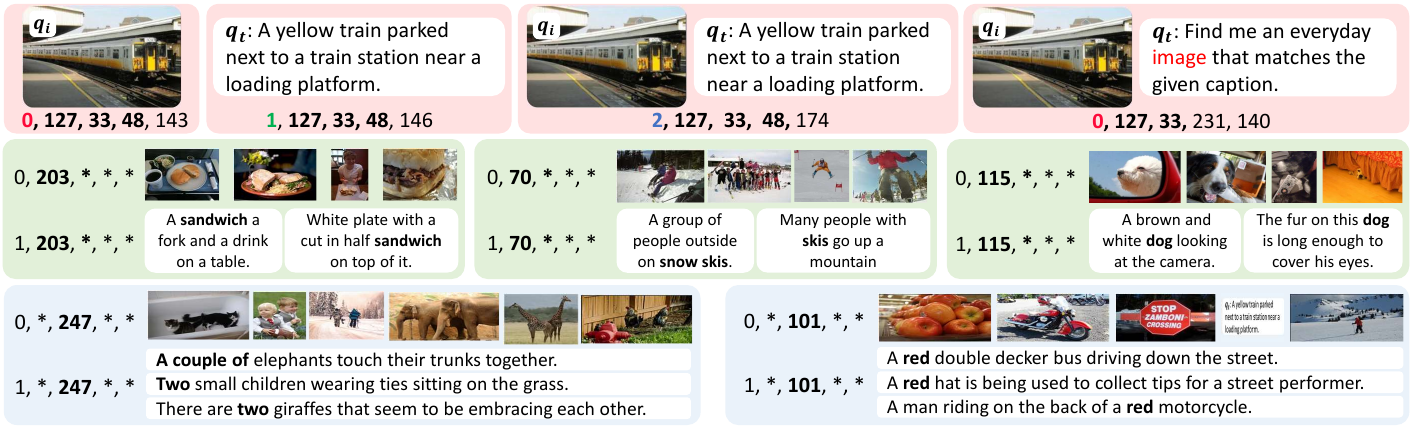}
\vspace*{-6mm}
\caption{\textbf{Examples of modality-decoupled semantic quantization.} For simplicity, we use a quantization scheme with five levels of codes, where each code (except the first) has a value of up to 256. The first code (top) indicates modality: 0 for image, 1 for text, and 2 for image-text pairs. If an instruction is provided, this code adapts to the modality specified by the instruction. The second code (middle) represents primary objects or dominant semantics shared across modalities, while the third code (bottom) captures key attributes of the main object, such as ``two” or ``red”, which are consistent across objects or data types. Beyond these levels, finer and additional information is incorporated to enrich the representation. This visualization is based on examples from the COCO dataset~\cite{Mscoco}.
}
\label{fig:code_visual}
\vspace*{-2mm}
\end{figure*}

\subsection{Modality-Decoupled Semantic Quantization}
\label{subsec:quantizaiton}
In generative retrieval, targets are represented as {discrete IDs} forming the output structure of the decoder model. Quantizing targets into these IDs is crucial, directly impacting retrieval performance. Unlike existing methods, GENIUS retrieves target data across modalities, and thus, it is essential to distinguish different modalities while accurately capturing semantic content. 

To this end, we propose a quantization method that represents \emph{modality} and \emph{semantic} information separately.
Our key idea is to provide an embedding space that captures both modality and semantic information using \emph{contrastive learning with queries including instructions} and to systematically separate these features through \emph{residual quantization} (RQ)~\cite{Lee_2022_CVPR, zeghidour2021soundstream, rajput2023recommender}. Leveraging the unique property of residual quantization allows us to produce structured code sequences, where modality is explicitly encoded at the first level and semantic details are progressively refined in subsequent levels.

\subsubsection{Fusion Module for Quantization Input} 
To facilitate effective quantization that captures both modality and semantics, we construct unified multimodal embeddings as inputs to the quantization. For this purpose, we introduce a lightweight, learnable module that combines image and text features into a unified representation. Inspired by previous work~\cite{clip4cir}, the fusion module is defined as:
\[
h(x,y) = \lambda \cdot x + (1-\lambda) \cdot y + \texttt{MLP}([x;y]),
\]
where $\texttt{MLP}([x; y])$ introduces additional bimodal information through a multi-layer perceptron (MLP) applied to the concatenation of $x$ and $y$. The balance parameter $\lambda$  is dynamically determined via another MLP with a sigmoid activation over the concatenated image-text features.
The fused query embedding is computed as
$\mathbf{z}_q = h\left(\psi_{\text{image}}(q_i), \psi_{\text{text}}(q_t, q_{\text{inst}}) \right)\in\mathbb{R}^{d}$ and the fused target embedding as
$\mathbf{z}_c = h\left(\psi_{\text{image}}(c_i) , \psi_{\text{text}}(c_t) \right)\in\mathbb{R}^{d}$. 
The fusion module is optimized with the quantization module by the objectives of the quantization process.

\subsubsection{Contrastive Learning with Instruction}
We construct an embedding space that integrates both modality and semantic information to prepare input embeddings for the modality-decoupled quantization. Using queries including instructions that specify the desired modality of the target, we apply a contrastive loss to align between these queries and their corresponding targets. This loss encourages data with the same semantics and modality to close together in the embedding space while pushing apart data that differ in either aspect. 
The contrastive loss is defined as: \begin{equation} 
\mathcal{L}_{\text{cl}} = -\log \frac{\exp\left( \left\langle \mathbf{z}_q,\ \mathbf{z}_{c^+} \right\rangle / \tau \right)}{\sum_{c' \in \mathcal{C}} \exp\left( \left\langle \mathbf{z}_q,\ \mathbf{z}_{c'} \right\rangle / \tau \right)}, \label{eq:loss_cl} 
\end{equation} where $\mathbf{z}_q$  and $\mathbf{z}_{c^+}$ is the query and the corresponding target embedding, $\mathcal{C}$ is the set of all candidate targets. Through this loss, clusters form in the embedding space, where modality-based groups naturally form due to the larger sample size within each modality, while semantically similar data are closely aligned within these clusters.

\subsubsection{Residual Quantization} 
Residual quantization (RQ)~\cite{Lee_2022_CVPR, zeghidour2021soundstream, rajput2023recommender} is a recursive process that approximates an embedding by quantizing its residuals at each level. 
This process enables a progressive information decomposition, allowing distinct levels to capture modality-specific and semantic elements separately.
The RQ process converts the embedding $\mathbf{z}$ into a sequence of discrete codes, represented as: 
\begin{equation}
    T := \mathcal{RQ}(\mathbf{z}) = (t_1, \dots , t_M),
\end{equation}
where $M$ is the number of quantization levels. 
Starting with the initial residual vector $r_0 = \mathbf{z}$, we perform quantization recursively. At each step $i$, we find the nearest neighbor within the $i$-th codebook $E_i = \{ \mathbf{e}_i^k \in\mathbb{R}^{d} \mid k = 1, \dots, K_i \}$, where $K_i$ is the size of the $i$-th codebook, selecting the closest code embedding $\mathbf{e}_{i}^{t_i}$ to the current residual vector:
\begin{equation}
t_i = \argmin_{k \in K_i } || \mathbf{r}_{i-1} - \mathbf{e}_i^{k} ||^2,
\end{equation}
and then update the residual for the next level:
\begin{equation}
\mathbf{r}_i = \mathbf{r}_{i-1} - \mathbf{e}_{i}^{t_i},
\end{equation}
The original embedding is approximated by summing the code embeddings up to level $M$, and we define this approximation as the quantized vector, $\hat{\mathbf{z}} = \sum_{i=1}^M \mathbf{e}^{t_i}_i$. 
Our key idea is to exploit the inherent property of residual quantization, where code embeddings at each level represent the residual information specific to that level. This property enables the progressive separation of information across levels. We utilize this property to distinguish modality and semantic information at each level. The first code in each ID explicitly represents modality, with a codebook of size $K_1 = 3$ to indicate images, text, and image-text pairs. Subsequent residuals exclude modality information, allowing the remaining levels to encode semantics solely in a coarse-to-fine manner.

\subsubsection{Training Objectives}
For training the codebooks and the fusion module $h$, we adopt three losses as follows. 
To ensure alignment between the assigned codes and the original residuals, we apply a residual quantization loss: 
\begin{equation}
\mathcal{L}_{\text{rq}} = \sum_{i=1}^M || \mathbf{r}_{i-1} - \text{sg}\left( \mathbf{e}_{i}^{t_i} \right) ||^2,
\label{eq:loss_ml}
\end{equation}
where $\text{sg}(\cdot)$ denotes the stop-gradient operator, preventing gradients from directly updating codebook entries.
Instead, they are updated via an exponential moving average (EMA)~\cite{razavi2019generating} over training steps to ensure stable updates.
In addition, to further reinforce semantic similarity in the quantized space, we introduce a mean squared error (MSE) loss between the quantized vector of the query and target as $
\mathcal{L}_{\text{mse}} = || \hat{\mathbf{z}}_q - \hat{\mathbf{z}}_c ||^2$, where $\hat{\mathbf{z}}_q$ and $\hat{\mathbf{z}}_c$ are the quantized query and target vectors, respectively.
The training loss is a linear combination of the three aforementioned losses:
\begin{equation} 
\mathcal{L}_{\text{combined}} = \mathcal{L}_{\text{cl}} + \beta \mathcal{L}_{\text{rq}} + \gamma \mathcal{L}_{\text{mse}},
\label{eq:rq_total_loss}
\end{equation}
where $\beta$ and $\gamma$ are weighting parameters.
Unlike prior methods focused on reconstructing original embeddings~\cite{li2024grace}, our optimization aims to encode contrastive relations into the codebook. As a result, the quantizer produces the initial code representing modality, as shown in Fig.~\ref{fig:code_visual}. The second code captures dominant semantics, while later codes add finer attributes, creating a structured representation that preserves rich, interpretable semantics and enhances retrieval performance across modalities.

\subsection{Autoregressive Decoder for Retrieval}
\label{subsec:decoder}
\subsubsection{Decoder Training}


The last step is to train an autoregressive decoder model that produces an ID of the target given a query. We adopt T5 decoder architecture~\cite{2020t5}, which generates the target ID autoregressively.
To condition the decoder on the query embedding, we employ a lightweight network with an MLP that maps the query embedding \(\mathbf{z}_q\) into \(N\) prefix embeddings, reshaping it as follows:
\begin{equation}
\mathbf{P}(\mathbf{z}_q) = \texttt{Reshape}(\texttt{MLP}(\mathbf{z}_q)) \in \mathbb{R}^{N \times d'},
\label{eq:prefix}
\end{equation}
where \(d'\) represents the hidden dimension of the decoder. These prefix embeddings $\mathbf{P}(\mathbf{z}_q)$ are fed to the decoder through cross-attention, enabling it to generate target IDs based on the semantic information embedded in the query.
The training loss for this generative model is a cross-entropy loss applied over the generated ID as follows:
\begin{equation}
\mathcal{L}_{\text{GR}}(\mathbf{P}(\mathbf{z}_q), T_c) = - \sum_{k=1}^{M} \log p\left( t^c_k \mid \mathbf{P}(\mathbf{z}_q), t^c_{<k} \right).
\end{equation}
This encourages the model to generate a target code sequence conditioned on the query, which can be considered mapping a query embedding to the target ID.


However, due to the inherently limited representation capacity in these discrete IDs compared to embeddings, the model may struggle to generalize effectively, particularly in scenarios with few query-target pairs for training. 
In text document generative retrieval, this challenge arises but is often addressed by generating diverse queries from documents using methods like Doc2Query~\cite{nogueira2019document, nogueira2019doc2query}; however, such methods are not feasible in multimodal retrieval.

\subsubsection{Query Augmentation via Interpolation}
To address the above issue, we propose a \textit{Query Augmentation} based on query-target interpolation. This technique enriches the training data by generating diverse augmented queries that remain semantically aligned with their target.
The interpolated query embedding \(\mathbf{z}'_q\) is computed as:
\begin{equation}
\mathbf{z}'_q = \mu \cdot \mathbf{z}_q + (1 - \mu) \cdot \mathbf{z}_c,
\label{eq:query_aug}
\end{equation}
where \(\mu\) is randomly sampled from a Beta distribution, \(\text{Beta}(\alpha, \alpha)\). The decoder is trained with the same cross-entropy loss with the augmented query, \(\mathcal{L}_{\text{GR}}(\mathbf{P}(\mathbf{z}'_q), T_c)\).
This strategy generates varied augmented queries, each maintaining relevance to the target, helping the decoder to learn a generalized mapping from query embeddings to target IDs. This makes the model more robust to variations in the query, improving its generalization.

\subsection{Inference}
\label{subsec:inference}
\paragraph{Constrained beam search.}
GENIUS retrieves relevant targets for inference by generating IDs based on a given query. To produce a ranked list of candidates, we use beam search, which explores multiple ID sequences and ranks them by the sum of the log probabilities for each level in the sequence. However, to prevent the risk of generating invalid IDs, we use \textit{constrained beam search}~\cite{de2020autoregressive} with a Trie structure~\cite{fredkin1960trie} that restricts the model to only valid prefixes matching actual test set IDs. The Trie is pre-constructed from all candidate IDs, allowing the decoder to ensure that generated IDs are valid. The time complexity for searching using Trie is $O(M)$, depending only on the length 
$M$ of the IDs, which can significantly enhance scalability.
\vspace{-4mm}
\paragraph{Embedding-based re-ranking.}
Despite this efficiency, generative retrieval with discrete IDs often lags behind embedding-based retrieval in performance due to the limitations of discrete representations, as observed in prior work~\cite{li2024grace}.
To address this, we present a re-ranking method: after predicting $B$ candidate IDs via beam search, we measure the similarity between the embeddings of these candidates and the query embedding. Since the number of comparisons is small, this method incurs negligible computational cost while greatly improving retrieval accuracy.

\section{Experiments}
\label{sec:experiments}

\begin{table*}[t]
\centering
\fontsize{7.5}{9}\selectfont  
\setlength{\tabcolsep}{3.5pt} 
\begin{tabularx}{\textwidth}{p{0.09\textwidth} |ccc|c|cc|ccc|c|cc|cc|cc}
\toprule
\multirow{3}{*}[-1mm]{\textbf{Method}} & \multicolumn{3}{c|}{$q_t$ $\to$ $c_i$} & {$q_t$ $\to$ $c_t$} & \multicolumn{2}{c|}{$q_t$ $\to$ ($c_i$, $c_t$)} & \multicolumn{3}{c|}{$q_i$ $\to$ $c_t$} & {$q_i$ $\to$ $c_i$} & \multicolumn{2}{c|}{($q_i$, $q_t$) $\to$ $c_t$} & \multicolumn{2}{c|}{($q_i$, $q_t$) $\to$ $c_i$} & \multicolumn{2}{c}{($q_i$, $q_t$) $\to$ ($c_i$, $c_t$)} \\
\cmidrule(lr){2-4} \cmidrule(lr){5-5} \cmidrule(lr){6-7} \cmidrule(lr){8-10} \cmidrule(lr){11-11} \cmidrule(lr){12-13} \cmidrule(lr){14-15} \cmidrule(lr){16-17} 
 & COCO & VN  & F200K & WebQA & EDIS & WebQA & COCO & VN & F200K & NIGHTS & OVEN & InfoS & FIQ & CIR & OVEN & InfoS \\
 & R@5 &R@5 &R@10 &R@5 &R@5 &R@5 &R@5 &R@5 &R@10 &R@5 &R@5 &R@5 &R@10 &R@5 &R@5 &R@5 \\
 
\midrule
\multicolumn{17}{c}{\textit{\textbf{Embedding-based Retrieval}}}  \\  [-0.4ex]
\midrule
{CLIP-SF}~\cite{uniir} & \textbf{81.1} & \textbf{42.6}  & 18.0 & \textbf{84.7} & \textbf{59.4} & 78.7  & \textbf{92.3} & \textbf{43.1}  & 18.3 & 32.0 & \textbf{45.5} & \textbf{27.9} & 24.4 & 44.6 & \textbf{67.6} & \textbf{48.9 } \\
{BLIP-FF}~\cite{uniir} & 79.7 & 23.4  & \textbf{26.1} & 80.0 & 50.9 & \textbf{79.8} & 89.9 & 22.8  & \textbf{28.9} & \textbf{33.0} & 41.0 & 22.4 & \textbf{29.2} & \textbf{52.2} & 55.8 & 33.0  \\
\midrule
\multicolumn{17}{c}{\textit{\textbf{Generative Retrieval}}}  \\  [-0.4ex]
\midrule
{IRGen}~\cite{zhang2023irgen} & 50.7 & --  & -- & -- & -- & -- & -- & -- & -- & -- & -- & -- & -- & -- & -- & --  \\
{GRACE}~\cite{li2024grace} & 39.5 & --  & -- & -- & -- & -- & -- & -- & -- & -- & -- & -- & -- & -- & -- & --  \\
\textbf{GENIUS}  
& 68.1 & 18.5  & 13.7 
& 32.5
& 37.0 & 49.7
& 83.2 & 18.7  & 12.8
& 8.2 
& 36.6 & 11.2
& 13.2 & 20.7
& 36.4 & 14.6 \\
\textbf{GENIUS$^{\mathcal{R}}$} 
& \textbf{78.0} & \textbf{27.4}  & \textbf{16.2} 
& \textbf{44.6} 
& \textbf{44.3} & \textbf{60.6} 
& \textbf{91.1} & \textbf{28.4}  & \textbf{16.3} 
& \textbf{30.2}
& \textbf{41.9} & \textbf{20.7} 
& \textbf{19.3} & \textbf{39.5} 
& \textbf{52.5} & \textbf{30.1}  \\
\bottomrule
\end{tabularx}
\vspace{-2mm}
\caption{\textbf{Task-specific Information Retrieval.} Performance of methods on the M-BEIR dataset,
retrieved from a task-specific pool. $\mathcal{R}$ denotes re-ranking using embedding vectors within the set of predicted candidates. Some datasets are denoted by abbreviations: VN--VisualNews, F200K--Fashion200K, InfoS--InfoSeek, and FIQ--FashionIQ.
}
\label{tab:results_local}
\vspace{-3mm}
\end{table*}

To evaluate the effectiveness of our generative universal retrieval framework, we conducted comprehensive experiments across various retrieval tasks and domains, comparing our model against state-of-the-art baselines in both embedding-based and generative retrieval paradigms.

\subsection{Dataset and metrics} \noindent{\textbf{Dataset.}} 
We use M-BEIR dataset~\cite{uniir}, a combination of multiple datasets. It includes datasets like MS-COCO~\cite{Mscoco} for image-caption retrieval, Fashion200K~\cite{han2017automatic} and FashionIQ~\cite{wu2021fashion} for fashion, VisualNews~\cite{liu2017visual} for news images, and NIGHTS~\cite{fu2023dreamsim} for image similarity. Complex retrieval tasks are addressed by OVEN~\cite{hu2023open}, EDIS~\cite{liu2023edis}, and CIRR~\cite{liu2021image}, with InfoSeek~\cite{chen2023can} and WebQA~\cite{chang2022webqa} for VQA-based retrieval. These datasets cover 8 multimodal tasks and have a total of 5.6 million candidates.

\noindent{\textbf{Evaluation metrics.}} 
Following prior work~\cite{uniir}, we report Recall@5 (R@5) as the main metric, using Recall@10 (R@10) for Fashion200K and FashionIQ.

\subsection{Implementation Details}
\noindent{\textbf{Network architectures.}} 
Following UniIR~\cite{uniir}, we use the pre-trained CLIP ViT-L/14 model~\cite{clip} as the vision and text encoder. For the decoder, we use T5-small~\cite{2020t5}, with hidden dimension \(d' = 512\), which is initialized randomly. 

\noindent{\textbf{Network optimization.}} 
Our model is optimized with AdamW, using a learning rate of \(1 \times 10^{-4}\) for both the RQ and decoder training. Residual quantization is trained for 20 epochs, while the decoder is trained for 30 epochs with cosine scheduling. We use a batch size of 256 for training. 

\noindent{\textbf{Hyperparameters.}} The contrastive learning temperature \(\tau\) in Eq.~\ref{eq:pt_contra} is set to 0.01. Parameters \(\beta\) and \(\gamma\) are both fixed at 100 Eq.~\ref{eq:rq_total_loss}, and \(\alpha\) parameter in Eq.~\ref{eq:query_aug} is set to 2. For the prefix embeddings in Eq.~\ref{eq:prefix}, we use a fixed length of 30.

\noindent{\textbf{Codebook configurations of RQ.}} 
Our default setting uses a codebook size of 4096 with 9 levels, except for the first codebook, which has a fixed size of 3. The codebook is initialized using $k$-means clustering on the first training batch.

\noindent{\textbf{Inference.}}
As described in Section \ref{subsec:inference}, we evaluate GENIUS in two ways: (i) constrained beam search and (ii) re-ranking beam search candidates based on their embeddings and that of query, both using a default beam size of 50 unless otherwise specified. The embedding-based methods are evaluated using nearest neighbor search by Faiss~\cite{douze2024faiss}.

\subsection{Baselines}
\noindent{\textbf{Training strategies.}} We evaluate models under two different training strategies: (i) \emph{single-task fine-tuning}, where models are independently trained and evaluated on each specific task, and (ii) \emph{unified instruction fine-tuning}, where models leverage multi-task learning with instructional guidance on M-BEIR~\cite{uniir}, enabling a single model to handle retrieval tasks across multiple domains and modalities.

\noindent{\textbf{Embedding-based retrieval baselines.}} We compare GENIUS with fine-tuned variants of CLIP~\cite{clip} and BLIP~\cite{li2022blip} proposed in UniIR~\cite{uniir}. These baselines employ two fusion strategies: {score-level fusion (SF)}, which fuses information at the output embedding level, and {feature-level fusion (FF)}, which uses transformers to achieve feature fusion. 

\noindent{\textbf{Generative retrieval baselines.}} We benchmark against GRACE~\cite{li2024grace} and IRGen~\cite{zhang2023irgen} which is originally for image-to-image retrieval, adapted for text-to-image retrieval by replacing image inputs with text, reported in~\cite{li2024revolutionizing}. Note that previous generative methods are designed for a single task.

\begin{table}[!t]
\setlength{\tabcolsep}{1pt}
\centering
\fontsize{7.5}{9}\selectfont
\scalebox{1.0}{ 
\begin{tabularx}{\columnwidth}{ 
    p{0.27\columnwidth} 
    p{0.16\columnwidth} 
    >{\centering\arraybackslash}X
    >{\centering\arraybackslash}X 
    >{\centering\arraybackslash}X 
    >{\centering\arraybackslash}X 
}
\toprule
\multicolumn{2}{l}{} & \multicolumn{2}{c}{\textbf{Embedding-based}} & \multicolumn{2}{c}{\textbf{Generative}}   \\
\cmidrule(lr){3-4} \cmidrule(lr){5-6}
\textbf{Task} & \textbf{Dataset} & ${\text{CLIP}_\text{SF}}$ & ${\text{BLIP}_\text{FF}}$ & \textbf{GENIUS}  & \textbf{GENIUS$^\mathcal{R}$}  \\
\midrule
\multirow{3}{*}{$q_t \to c_i$} 
 & VisualNews & \textbf{42.6}& 23.0 & 18.5 & \textbf{27.3}\\
 & MSCOCO & \textbf{77.9} & 75.6 & 55.1 &\textbf{68.0} \\
 & Fashion200K & 17.8 & \textbf{25.4} & 13.7&\textbf{16.2} \\
\midrule
\multirow{1}{*}{$q_t \to c_t$} 
 & WebQA & \textbf{84.7}& 79.5 & 31.1 & \textbf{42.9}\\
\midrule
\multirow{2}{*}{$q_t \to (c_i, c_t)$} 
 & EDIS &\textbf{59.4} & 50.3 & 36.6& \textbf{44.1}\ \\
 & WebQA & 78.8 &\textbf{ 79.7} & 49.0 & \textbf{59.7} \\
\midrule
\multirow{3}{*}{$q_i \to c_t$} 
 & VisualNews &\textbf{42.8}& 21.1 & 18.4 & \textbf{26.8} \\
 & MSCOCO & \textbf{92.3}& 88.8 &82.7 & \textbf{90.6}\\
 & Fashion200K & 17.9 & \textbf{27.6} & 12.8 & \textbf{16.2} \\
\midrule
\multirow{1}{*}{$q_i \to c_i$} 
 & NIGHTS &\textbf{33.4} & 33.0 & 8.1& \textbf{30.2}  \\
\midrule
\multirow{2}{*}{$(q_i, q_t) \to c_t$} 
 & OVEN &\textbf{39.2} & 34.7 & 34.6 & \textbf{38.0}\\
 & InfoSeek & \textbf{24.0} & 19.7 & 10.4 & \textbf{18.0} \\
\midrule
\multirow{2}{*}{$(q_i, q_t) \to c_i$} 
 & FashionIQ& 26.2 & \textbf{28.5} &  13.1&\textbf{19.2} \\
 & CIRR  & 43.0 & \textbf{51.4}&20.1 & \textbf{38.3} \\
\midrule
\multirow{2}{*}{$(q_i, q_t) \to (c_i, c_t)$} 
 & OVEN &\textbf{60.2} & 57.8& 36.5& \textbf{48.6} \\
 & InfoSeek & \textbf{44.6} & 27.7 & 14.2 & \textbf{28.6} \\
\midrule
\textbf{Average} & & \textbf{48.9}& 45.5  & 27.6 & \textbf{38.3}  \\
\bottomrule
\end{tabularx}
}
\vspace{-2mm}
\caption{\textbf{Universal Information Retrieval.} 
Recall@5 results of methods except Fashion200K and FashionIQ, where Recall@10 is reported. Retrieval is performed from a global pool spanning diverse modalities. $\mathcal{R}$ denotes re-ranking using embedding vectors within the set of predicted candidates.}
\vspace{-5mm}
\label{tab:results_full}
\end{table}

\subsection{Experimental Results}
We evaluate multimodal retrieval models in three scenarios: (i) \emph{task-specific information retrieval}, using original datasets to ensure a fair comparison with single-task methods; (ii) \emph{universal information retrieval}, leveraging the full M-BEIR candidate pool of 5.6M items to assess models' capability in instruction-following and cross-modal retrieval tasks, a setting unsupported by existing generative approaches; and (iii) \emph{text-to-image generative retrieval}, evaluated on standard generative retrieval benchmarks (Flickr30K and MS-COCO), with models trained and evaluated separately on each dataset.

\noindent{\textbf{Task-specific information retrieval.}}
In Table~\ref{tab:results_local}, GENIUS is compared against embedding-based retrieval methods (CLIP-SF and BLIP-FF) and existing generative retrieval baselines (GRACE and IRGen) on various datasets from M-BEIR. Generative retrieval methods show significantly lower performance compared to embedding-based approaches, even on single-task retrieval. Notably, GENIUS significantly outperforms previous generative methods on COCO text-to-image retrieval by 28.6 points in R@5, substantially narrowing the gap with embedding-based methods. GENIUS demonstrates competitive performance across multiple datasets, with embedding-based re-ranking further enhancing its effectiveness, enabling it to surpass BLIP-FF in several tasks.  However, GENIUS underperforms on knowledge-intensive retrieval tasks (\eg, WebQA, InfoSeek) compared to embedding-based retrieval. This limitation is likely due to the inherent capacity of discrete IDs, which should be addressed in future research.

\noindent{\textbf{Universal information retrieval.}} Table~\ref{tab:results_full} presents results for a range of retrieval tasks on entire candidates in M-BEIR dataset. Unlike prior settings, this universal scenario requires models to identify target modalities precisely based solely on given instructions. GENIUS demonstrates competitive performance and versatility across multimodal tasks, though it typically achieves lower results than embedding-based retrieval baselines. 

\noindent{\textbf{Text-to-image generative retrieval.}} Table~\ref{suptab:flickr_coco} compares GENIUS against recent generative retrieval models on the Flickr30K~\cite{Flickr30k_b} and MS-COCO~\cite{Mscoco} datasets. GENIUS significantly outperforms existing generative baselines such as GRACE and IRGen, showing substantial improvements across all metrics on both datasets. Further performance gains are achieved through embedding-based re-ranking, which yields state-of-the-art results in generative retrieval.


\begin{table}[!t]
\centering
\setlength{\tabcolsep}{4pt}
\fontsize{7.5}{9}\selectfont
\scalebox{1.0}{
\begin{tabularx}{\columnwidth}{
    p{0.2\columnwidth} |
    >{\centering\arraybackslash}X
    >{\centering\arraybackslash}X 
    >{\centering\arraybackslash}X |
    >{\centering\arraybackslash}X
    >{\centering\arraybackslash}X
    >{\centering\arraybackslash}X
}
\toprule
\multirow{2}{*}[-1mm]{\textbf{Method}} & \multicolumn{3}{c}{\textbf{Flickr30K}} & \multicolumn{3}{c}{\textbf{COCO}} \\ [-0.4ex]
\cmidrule(lr){2-4}  \cmidrule(lr){5-7} & R@1 & R@5 & R@10 & R@1 & R@5 & R@10 \\ [-0.4ex]
\midrule
GRACE~\cite{li2024grace} & 37.4 & 59.5 & 66.2 & 16.7 & 39.2 & 50.3 \\
IRGen~\cite{zhang2023irgen} & 49.0 & 68.9 & 72.5 & 29.6 & 50.7 & 56.3 \\
\textbf{GENIUS} & 60.6 & 84.0 & 90.5 & 40.1 & 66.2& 75.8\\
\textbf{GENIUS}$^{\mathcal{R}}$  & \textbf{74.1} & \textbf{92.0} & \textbf{94.8} & \textbf{46.1} & \textbf{74.0}& \textbf{82.7}\\

\bottomrule
\end{tabularx}
}
\vspace{-2mm}
\caption{
Text-to-image retrieval performance comparison on standard generative retrieval benchmark (Flickr30K and MS-COCO). $\mathcal{R}$ denotes re-ranking. Note that all models, including GENIUS, are trained and evaluated separately on each dataset.
}
\label{suptab:flickr_coco}
\vspace{-1mm}
\end{table}

\begin{table}[t!]
\centering
\fontsize{7.5}{9}\selectfont
\setlength{\tabcolsep}{4pt}
\scalebox{0.95}{
\begin{tabularx}{0.499\textwidth}{ 
   p{0.159\textwidth} |
   >{\centering\arraybackslash}p{0.04\textwidth} |
   >{\centering\arraybackslash}p{0.04\textwidth} |
   >{\centering\arraybackslash}p{0.04\textwidth} |
   >{\centering\arraybackslash}p{0.058\textwidth} |
   >{\centering\arraybackslash}p{0.058\textwidth}
   } 
\toprule
 & \multicolumn{2}{c}{\textbf{COCO}} & \multicolumn{2}{c}{\textbf{WebQA}} & \textbf{CIRR} \\ [-0.4ex] 
\cmidrule(lr){2-3} \cmidrule(lr){4-5} \cmidrule(lr){6-6} 
 \textbf{Method} & \fontsize{7}{8}\selectfont{T $\to$ I} & \fontsize{7}{8}\selectfont{I $\to$ T} & \fontsize{7}{8}\selectfont{T $\to$ T} & \fontsize{7}{8}\selectfont{T $\to$ (I,T)} & \fontsize{7}{8}\selectfont{(I,T) $\to$ I} 
  \\ [-0.4ex] \midrule
GENIUS & \textbf{55.4} & {82.7}& \textbf{28.3}&\textbf{ 47.1}&20.5 \\
w/o Modality-decoupled& 20.2&73.2 &25.9&34.3 &18.3\\
w/o Query augmentation  & 47.8&67.7 &19.6&38.8 &11.7\\ 
w/o $\mathcal{L}_{\text{cl}}$ in Eq.~\ref{eq:rq_total_loss} & 0.0&0.1 &0.0&0.0 &0.0 \\ 
w/o $\mathcal{L}_{\text{mse}}$ in Eq.~\ref{eq:rq_total_loss} & 45.5&\textbf{83.1}&27.1 &35.2 & \textbf{21.6} \\ 
\bottomrule
\end{tabularx}
}
\vspace{-2mm}
\caption{Ablation study for key components of GENIUS (universal information retrieval, R@5) with 30 beams. I and T denote image and text modalities, respectively, and (I,T) is image-text pair.}
\label{tab:main_compnent}
\vspace{-3mm}
\end{table}

\subsection{Analysis}
\noindent{\textbf{Ablation study on key components.}}
Table~\ref{tab:main_compnent} presents an ablation study of key components under retrieval from a global pool. Removing modality-decoupling severely harms modality discrimination, notably in COCO text-to-image retrieval.  Excluding query augmentation leads to decreased accuracy, highlighting its contribution to robustness. The contrastive loss ($\mathcal{L}_{\text{cl}}$) is crucial for aligning modality-decoupled representations; without it, query and target features become misaligned, leading to near-zero performance. Excluding MSE loss ($\mathcal{L}_{\text{mse}}$) weakens alignment in the codespace, reducing performance in certain datasets.

\begin{figure} [!t]
\centering
\vspace{-2mm}
\includegraphics[width = 0.65 \columnwidth]{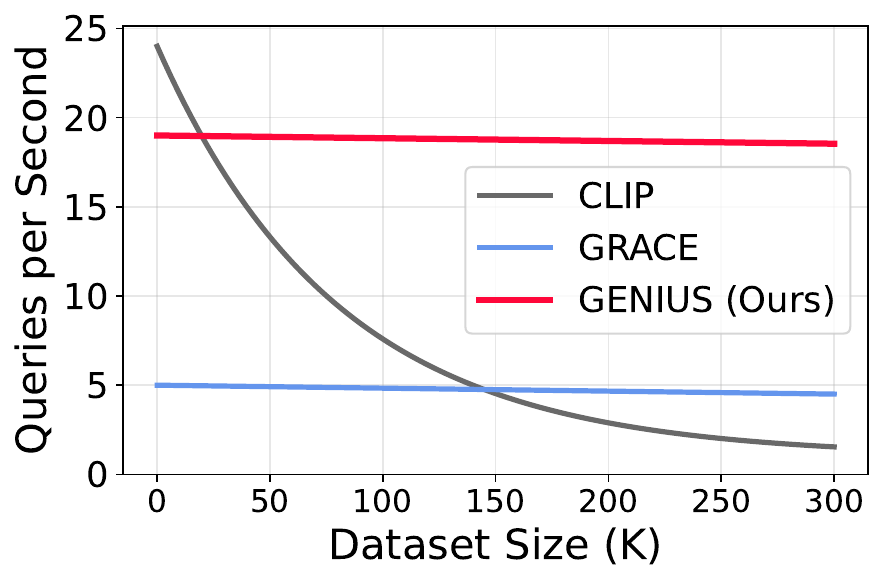}
\vspace{-3mm}
\caption{Efficiency in processed queries per second across varying dataset sizes, measured with a single RTX3090 GPU.
}
\label{fig:Efficiency}
\vspace{-1mm}
\end{figure}

\noindent{\textbf{Analysis on efficiency.}} 
We compare retrieval efficiency between embedding-based (CLIP) and generative methods (GRACE, GENIUS) by measuring queries per second, as shown in Fig.~\ref{fig:Efficiency}. For a fair comparison with GRACE, we use text queries with image candidates.  As the candidate dataset size increases, the efficiency of CLIP declines due to the growing cost of the nearest neighbor search, while generative methods remain nearly constant. GENIUS is a lightweight equipped with a T5-small decoder and CLIP encoder, and thus achieves roughly 4 times higher efficiency than GRACE with Flamingo-3B model~\cite{alayrac2022flamingo}. The efficiency advantage becomes more significant as the dataset grows, maintaining high retrieval speed at scale without the expensive index building typical in embedding-based methods.

\noindent{\textbf{Codebook configuration.}}
Table~\ref{tab:codebook} shows that larger codebook sizes and higher levels generally increase expressive power, and thus improve performance, especially in knowledge-intensive tasks such as WebQA. However, excessively large codebooks can disperse clusters, weakening representations in some datasets. This highlights the need to balance codebook size according to dataset characteristics.

\begin{table}[t!]
\centering
\fontsize{7.5}{8.5}\selectfont
\scalebox{0.95}{
\begin{tabularx}{0.499\textwidth}{ 
   p{0.12\textwidth} |
   >{\centering\arraybackslash}p{0.04\textwidth} |
   >{\centering\arraybackslash}p{0.04\textwidth} |
   >{\centering\arraybackslash}p{0.04\textwidth} |
   >{\centering\arraybackslash}p{0.058\textwidth} |
   >{\centering\arraybackslash}p{0.058\textwidth}
   } 
\toprule
 & \multicolumn{2}{c}{\textbf{COCO}} & \multicolumn{2}{c}{\textbf{WebQA}} & \textbf{CIRR} \\ [-0.4ex] 
\cmidrule(lr){2-3} \cmidrule(lr){4-5} \cmidrule(lr){6-6} 
\textbf{$K \times M$} & \fontsize{7}{8}\selectfont{T $\to$ I} & \fontsize{7}{8}\selectfont{I $\to$ T} & \fontsize{7}{8}\selectfont{T $\to$ T} & \fontsize{7}{8}\selectfont{T $\to$ (I,T)} & \fontsize{7}{8}\selectfont{(I,T) $\to$ I} 
  \\ [-0.4ex]  \midrule
\underline{4096}$\times$ \underline{9}~(Default) & {65.3} & \textbf{83.4} & \textbf{28.8}&\textbf{ 47.4 }&21.0 \\ \midrule
\underline{4096}$\times$ 7 & 65.2 & 82.9 & 25.3 & 40.8 & 23.1 \\
\underline{4096}$\times$ 5 & 62.4&81.5 &17.3 & 33.1&20.4 \\ \midrule
1024$\times$ \underline{9} & \textbf{66.4} & 82.0 &24.7 &39.4 & \textbf{24.5}\\
256$\times$ \underline{9} & 61.2& 76.6& 18.3& 33.5&18.3 \\ \midrule
1024$\times$ 7 & 64.3 &82.2 & 24.6 &42.7 &16.4 \\
256$\times$ 5 & 53.4 & 72.4 & 9.7 &22.8 &13.0 \\
\bottomrule
\end{tabularx}
}
\vspace{-2mm}
\caption{Ablation over codebook size $K$ (except for the first level) and code level $M$ (task-specific information retrieval, R@5) with 30 beams. The default codebook size and level are underlined.}
\label{tab:codebook}
\vspace{-3mm}
\end{table}


\section{Conclusion}
\label{sec:conclusion}
We have introduced GENIUS, a universal generative retrieval framework that addresses the limitations of existing generative models by handling diverse tasks across modalities. Leveraging a novel modality-decoupled quantization technique for ID generation, GENIUS ensures consistent semantic information across modalities. Our query augmentation enhances generalization through diverse query-target mappings. 
Experiments show that GENIUS outperforms prior generative methods and narrows the performance gap with embedding-based methods across benchmarks. Moreover, GENIUS sustains high retrieval speed, laying the groundwork for scalable multimodal search.


\section*{Acknowledgements}{
Part of this work was done while Sungyeon Kim was an intern at Amazon. Sungyeon Kim and Suha Kwak were supported by NRF grants (RS-2021-NR059830--30\%, RS-2022-II220290--30\%, RS-2022-II220926--30\%) and IITP grants (RS-2019-II191906--10\%, AI Graduate School - POSTECH) funded by Ministry of Science and ICT, Korea.
}

{\small
\bibliographystyle{ieeenat_fullname}
\bibliography{cvlab_kwak}

\begin{thebibliography}{64}
\providecommand{\natexlab}[1]{#1}
\providecommand{\url}[1]{\texttt{#1}}
\expandafter\ifx\csname urlstyle\endcsname\relax
  \providecommand{\doi}[1]{doi: #1}\else
  \providecommand{\doi}{doi: \begingroup \urlstyle{rm}\Url}\fi

\bibitem[Alayrac et~al.(2022)Alayrac, Donahue, Luc, Miech, Barr, Hasson, Lenc, Mensch, Millican, Reynolds, et~al.]{alayrac2022flamingo}
Jean-Baptiste Alayrac, Jeff Donahue, Pauline Luc, Antoine Miech, Iain Barr, Yana Hasson, Karel Lenc, Arthur Mensch, Katherine Millican, Malcolm Reynolds, et~al.
\newblock Flamingo: a visual language model for few-shot learning.
\newblock \emph{Proc. Neural Information Processing Systems (NeurIPS)}, 2022.

\bibitem[Bai et~al.(2023)Bai, Xu, Liu, Khan, Khan, Zuo, Goh, and Feng]{bai2023sentence}
Yang Bai, Xinxing Xu, Yong Liu, Salman Khan, Fahad Khan, Wangmeng Zuo, Rick Siow~Mong Goh, and Chun-Mei Feng.
\newblock Sentence-level prompts benefit composed image retrieval.
\newblock In \emph{Proc. International Conference on Learning Representations (ICLR)}, 2023.

\bibitem[Baldrati et~al.(2023)Baldrati, Bertini, Uricchio, and Bimbo]{clip4cir}
Alberto Baldrati, Marco Bertini, Tiberio Uricchio, and Alberto~Del Bimbo.
\newblock Composed image retrieval using contrastive learning and task-oriented clip-based features.
\newblock \emph{ACM Transactions on Multimedia Computing, Communications and Applications}, 2023.

\bibitem[Bevilacqua et~al.(2022)Bevilacqua, Ottaviano, Lewis, Yih, Riedel, and Petroni]{bevilacqua2022autoregressive}
Michele Bevilacqua, Giuseppe Ottaviano, Patrick Lewis, Scott Yih, Sebastian Riedel, and Fabio Petroni.
\newblock Autoregressive search engines: Generating substrings as document identifiers.
\newblock \emph{Proc. Neural Information Processing Systems (NeurIPS)}, 2022.

\bibitem[Chang et~al.(2022)Chang, Narang, Suzuki, Cao, Gao, and Bisk]{chang2022webqa}
Yingshan Chang, Mridu Narang, Hisami Suzuki, Guihong Cao, Jianfeng Gao, and Yonatan Bisk.
\newblock Webqa: Multihop and multimodal qa.
\newblock In \emph{Proc. IEEE Conference on Computer Vision and Pattern Recognition (CVPR)}, 2022.

\bibitem[Changpinyo et~al.(2021)Changpinyo, Pont-Tuset, Ferrari, and Soricut]{changpinyo2021telling}
Soravit Changpinyo, Jordi Pont-Tuset, Vittorio Ferrari, and Radu Soricut.
\newblock Telling the what while pointing to the where: Multimodal queries for image retrieval.
\newblock In \emph{Proc. IEEE Conference on Computer Vision and Pattern Recognition (CVPR)}, 2021.

\bibitem[Chen et~al.(2023)Chen, Hu, Luan, Sun, Changpinyo, Ritter, and Chang]{chen2023can}
Yang Chen, Hexiang Hu, Yi Luan, Haitian Sun, Soravit Changpinyo, Alan Ritter, and Ming-Wei Chang.
\newblock Can pre-trained vision and language models answer visual information-seeking questions?
\newblock In \emph{Proceedings of the 2023 Conference on Empirical Methods in Natural Language Processing}, 2023.

\bibitem[De~Cao et~al.(2020)De~Cao, Izacard, Riedel, and Petroni]{de2020autoregressive}
N De~Cao, G Izacard, S Riedel, and F Petroni.
\newblock Autoregressive entity retrieval.
\newblock In \emph{ICLR 2021-9th International Conference on Learning Representations}. ICLR, 2020.

\bibitem[Douze et~al.(2024)Douze, Guzhva, Deng, Johnson, Szilvasy, Mazaré, Lomeli, Hosseini, and Jégou]{douze2024faiss}
Matthijs Douze, Alexandr Guzhva, Chengqi Deng, Jeff Johnson, Gergely Szilvasy, Pierre-Emmanuel Mazaré, Maria Lomeli, Lucas Hosseini, and Hervé Jégou.
\newblock The faiss library.
\newblock \emph{arXiv preprint arXiv:2401.08281}, 2024.

\bibitem[Du et~al.(2024)Du, Xiu, and Tanaka-Ishii]{du2024bottleneck}
Xin Du, Lixin Xiu, and Kumiko Tanaka-Ishii.
\newblock Bottleneck-minimal indexing for generative document retrieval.
\newblock In \emph{Forty-first International Conference on Machine Learning}, 2024.

\bibitem[Formal et~al.(2021)Formal, Piwowarski, and Clinchant]{formal2021splade}
Thibault Formal, Benjamin Piwowarski, and St{\'e}phane Clinchant.
\newblock Splade: Sparse lexical and expansion model for first stage ranking.
\newblock In \emph{Proceedings of the 44th International ACM SIGIR Conference on Research and Development in Information Retrieval}, 2021.

\bibitem[Fredkin(1960)]{fredkin1960trie}
Edward Fredkin.
\newblock Trie memory.
\newblock \emph{Communications of the ACM}, 1960.

\bibitem[Fu et~al.(2023)Fu, Tamir, Sundaram, Chai, Zhang, Dekel, and Isola]{fu2023dreamsim}
Stephanie Fu, Netanel~Yakir Tamir, Shobhita Sundaram, Lucy Chai, Richard Zhang, Tali Dekel, and Phillip Isola.
\newblock Dreamsim: Learning new dimensions of human visual similarity using synthetic data.
\newblock In \emph{Thirty-seventh Conference on Neural Information Processing Systems}, 2023.

\bibitem[Girdhar et~al.(2023)Girdhar, El-Nouby, Liu, Singh, Alwala, Joulin, and Misra]{girdhar2023imagebind}
Rohit Girdhar, Alaaeldin El-Nouby, Zhuang Liu, Mannat Singh, Kalyan~Vasudev Alwala, Armand Joulin, and Ishan Misra.
\newblock Imagebind: One embedding space to bind them all.
\newblock In \emph{Proc. IEEE Conference on Computer Vision and Pattern Recognition (CVPR)}, 2023.

\bibitem[Han et~al.(2017)Han, Wu, Huang, Zhang, Zhu, Li, Zhao, and Davis]{han2017automatic}
Xintong Han, Zuxuan Wu, Phoenix~X Huang, Xiao Zhang, Menglong Zhu, Yuan Li, Yang Zhao, and Larry~S Davis.
\newblock Automatic spatially-aware fashion concept discovery.
\newblock In \emph{Proc. IEEE International Conference on Computer Vision (ICCV)}, 2017.

\bibitem[Hu et~al.(2023)Hu, Luan, Chen, Khandelwal, Joshi, Lee, Toutanova, and Chang]{hu2023open}
Hexiang Hu, Yi Luan, Yang Chen, Urvashi Khandelwal, Mandar Joshi, Kenton Lee, Kristina Toutanova, and Ming-Wei Chang.
\newblock Open-domain visual entity recognition: Towards recognizing millions of wikipedia entities.
\newblock In \emph{Proc. IEEE Conference on Computer Vision and Pattern Recognition (CVPR)}, 2023.

\bibitem[Jia et~al.(2021)Jia, Yang, Xia, Chen, Parekh, Pham, Le, Sung, Li, and Duerig]{align}
Chao Jia, Yinfei Yang, Ye Xia, Yi-Ting Chen, Zarana Parekh, Hieu Pham, Quoc Le, Yun-Hsuan Sung, Zhen Li, and Tom Duerig.
\newblock Scaling up visual and vision-language representation learning with noisy text supervision.
\newblock In \emph{Proc. International Conference on Machine Learning (ICML)}, 2021.

\bibitem[Jin et~al.(2024)Jin, Zeng, Wang, Chen, Wei, Li, Wang, Li, Li, Lu, Wang, Han, and Tang]{jin2023language}
Bowen Jin, Hansi Zeng, Guoyin Wang, Xiusi Chen, Tianxin Wei, Ruirui Li, Zhengyang Wang, Zheng Li, Yang Li, Hanqing Lu, Suhang Wang, Jiawei Han, and Xianfeng Tang.
\newblock Language models as semantic indexers.
\newblock In \emph{Forty-first International Conference on Machine Learning}, 2024.

\bibitem[Karpathy()]{Karpathy_mscoco_split}
Andrej Karpathy.
\newblock Neuraltalk2.
\newblock https://github.com/karpathy/neuraltalk2.

\bibitem[Kim et~al.(2023)Kim, Kim, and Kwak]{kim2023improving}
Dongwon Kim, Namyup Kim, and Suha Kwak.
\newblock Improving cross-modal retrieval with set of diverse embeddings.
\newblock In \emph{Proc. IEEE Conference on Computer Vision and Pattern Recognition (CVPR)}, 2023.

\bibitem[Kim et~al.(2020)Kim, Kim, Cho, and Kwak]{kim2020proxy}
Sungyeon Kim, Dongwon Kim, Minsu Cho, and Suha Kwak.
\newblock Proxy anchor loss for deep metric learning.
\newblock In \emph{Proc. IEEE Conference on Computer Vision and Pattern Recognition (CVPR)}, 2020.

\bibitem[Lee et~al.(2022)Lee, Kim, Kim, Cho, and Han]{Lee_2022_CVPR}
Doyup Lee, Chiheon Kim, Saehoon Kim, Minsu Cho, and Wook-Shin Han.
\newblock Autoregressive image generation using residual quantization.
\newblock In \emph{Proc. IEEE Conference on Computer Vision and Pattern Recognition (CVPR)}, 2022.

\bibitem[Li et~al.(2020)Li, Duan, Fang, Gong, and Jiang]{li2020unicoder}
Gen Li, Nan Duan, Yuejian Fang, Ming Gong, and Daxin Jiang.
\newblock Unicoder-vl: A universal encoder for vision and language by cross-modal pre-training.
\newblock In \emph{Proc. AAAI Conference on Artificial Intelligence (AAAI)}, 2020.

\bibitem[Li et~al.(2022)Li, Li, Xiong, and Hoi]{li2022blip}
Junnan Li, Dongxu Li, Caiming Xiong, and Steven Hoi.
\newblock Blip: Bootstrapping language-image pre-training for unified vision-language understanding and generation.
\newblock In \emph{International conference on machine learning}, pages 12888--12900. PMLR, 2022.

\bibitem[Li et~al.(2023)Li, Li, Savarese, and Hoi]{li2023blip2}
Junnan Li, Dongxu Li, Silvio Savarese, and Steven Hoi.
\newblock Blip-2: Bootstrapping language-image pre-training with frozen image encoders and large language models.
\newblock In \emph{Proc. International Conference on Machine Learning (ICML)}, 2023.

\bibitem[Li et~al.(2024{\natexlab{a}})Li, Cai, Wang, Qu, Wei, Li, Nie, and Chua]{li2024revolutionizing}
Yongqi Li, Hongru Cai, Wenjie Wang, Leigang Qu, Yinwei Wei, Wenjie Li, Liqiang Nie, and Tat-Seng Chua.
\newblock Revolutionizing text-to-image retrieval as autoregressive token-to-voken generation.
\newblock \emph{arXiv preprint arXiv:2407.17274}, 2024{\natexlab{a}}.

\bibitem[Li et~al.(2024{\natexlab{b}})Li, Wang, Qu, Nie, Li, and Chua]{li2024grace}
Yongqi Li, Wenjie Wang, Leigang Qu, Liqiang Nie, Wenjie Li, and Tat-Seng Chua.
\newblock Generative cross-modal retrieval: Memorizing images in multimodal language models for retrieval and beyond.
\newblock In \emph{Annual Meeting of the Association for Computational Linguistics (ACL)}, 2024{\natexlab{b}}.

\bibitem[Lin et~al.(2014)Lin, Maire, Belongie, Hays, Perona, Ramanan, Doll{\'a}r, and Zitnick]{Mscoco}
Tsung-Yi Lin, Michael Maire, Serge Belongie, James Hays, Pietro Perona, Deva Ramanan, Piotr Doll{\'a}r, and C~Lawrence Zitnick.
\newblock {Microsoft COCO:} common objects in context.
\newblock In \emph{Proc. European Conference on Computer Vision (ECCV)}, 2014.

\bibitem[Liu et~al.(2010)Liu, Wang, Shi, Huang, Lin, Shen, Yang, Lu, and Yuille]{liu2017visual}
Fuwen Liu, Xiaojie Wang, Jianping Shi, Alan~L Huang, Zhe Lin, Xiaohui Shen, Jimei Yang, Xin Lu, and Alan Yuille.
\newblock Visual news: Benchmark and challenges in news image captioning.
\newblock In \emph{Proc. IEEE Conference on Computer Vision and Pattern Recognition (CVPR)}, 2010.

\bibitem[Liu et~al.(2023{\natexlab{a}})Liu, Feng, Fu, Chen, and Wang]{liu2023edis}
Siqi Liu, Weixi Feng, Tsu-jui Fu, Wenhu Chen, and William~Yang Wang.
\newblock Edis: Entity-driven image search over multimodal web content.
\newblock In \emph{Proceedings of the 2023 Conference on Empirical Methods in Natural Language Processing}, pages 4877--4894, 2023{\natexlab{a}}.

\bibitem[Liu et~al.(2021)Liu, Rodriguez-Opazo, Teney, and Gould]{liu2021image}
Zheyuan Liu, Cristian Rodriguez-Opazo, Damien Teney, and Stephen Gould.
\newblock Image retrieval on real-life images with pre-trained vision-and-language models.
\newblock In \emph{Proc. IEEE International Conference on Computer Vision (ICCV)}, 2021.

\bibitem[Liu et~al.(2023{\natexlab{b}})Liu, Xiong, Lv, Liu, and Yu]{liu2022universal}
Zhenghao Liu, Chenyan Xiong, Yuanhuiyi Lv, Zhiyuan Liu, and Ge Yu.
\newblock Universal vision-language dense retrieval: Learning a unified representation space for multi-modal retrieval.
\newblock In \emph{The Eleventh International Conference on Learning Representations (ICLR)}, 2023{\natexlab{b}}.

\bibitem[Luo et~al.(2023)Luo, Fang, Gokhale, Yang, and Baral]{luo2023end}
Man Luo, Zhiyuan Fang, Tejas Gokhale, Yezhou Yang, and Chitta Baral.
\newblock End-to-end knowledge retrieval with multi-modal queries.
\newblock In \emph{61st Annual Meeting of the Association for Computational Linguistics, ACL 2023}, pages 8573--8589. Association for Computational Linguistics (ACL), 2023.

\bibitem[Malkov and Yashunin(2018)]{malkov2018efficient}
Yu~A Malkov and Dmitry~A Yashunin.
\newblock Efficient and robust approximate nearest neighbor search using hierarchical navigable small world graphs.
\newblock \emph{IEEE Transactions on Pattern Analysis and Machine Intelligence (TPAMI)}, 2018.

\bibitem[Manning et~al.(2008)Manning, Raghavan, and Sch\"{u}tze]{manning2009introduction}
Christopher~D. Manning, Prabhakar Raghavan, and Hinrich Sch\"{u}tze.
\newblock \emph{Introduction to Information Retrieval}.
\newblock Cambridge University Press, USA, 2008.

\bibitem[McInnes et~al.(2018)McInnes, Healy, Saul, and Grossberger]{mcinnes2018umap}
Leland McInnes, John Healy, Nathaniel Saul, and Lukas Grossberger.
\newblock Umap: Uniform manifold approximation and projection.
\newblock \emph{The Journal of Open Source Software}, 2018.

\bibitem[Mehta et~al.(2023)Mehta, Gupta, Tay, Dehghani, Tran, Rao, Najork, Strubell, and Metzler]{mehta2022dsi++}
Sanket~Vaibhav Mehta, Jai Gupta, Yi Tay, Mostafa Dehghani, Vinh~Q Tran, Jinfeng Rao, Marc Najork, Emma Strubell, and Donald Metzler.
\newblock Dsi++: Updating transformer memory with new documents.
\newblock In \emph{Proceedings of the 2023 Conference on Empirical Methods in Natural Language Processing}, pages 8198--8213, 2023.

\bibitem[Movshovitz-Attias et~al.(2017)Movshovitz-Attias, Toshev, Leung, Ioffe, and Singh]{movshovitz2017no}
Yair Movshovitz-Attias, Alexander Toshev, Thomas~K Leung, Sergey Ioffe, and Saurabh Singh.
\newblock No fuss distance metric learning using proxies.
\newblock In \emph{Proc. IEEE International Conference on Computer Vision (ICCV)}, 2017.

\bibitem[Nguyen and Yates(2023)]{nguyen2023generative}
Thong Nguyen and Andrew Yates.
\newblock Generative retrieval as dense retrieval.
\newblock \emph{arXiv preprint arXiv:2306.11397}, 2023.

\bibitem[Nogueira and Lin(2019)]{nogueira2019doc2query}
Rodrigo Nogueira and Jimmy Lin.
\newblock From doc2query to doctttttquery.
\newblock \emph{Online preprint}, 2019.

\bibitem[Nogueira et~al.(2019)Nogueira, Yang, Lin, and Cho]{nogueira2019document}
Rodrigo Nogueira, Wei Yang, Jimmy Lin, and Kyunghyun Cho.
\newblock Document expansion by query prediction.
\newblock \emph{arXiv preprint arXiv:1904.08375}, 2019.

\bibitem[Radford et~al.(2021)Radford, Kim, Hallacy, Ramesh, Goh, Agarwal, Sastry, Askell, Mishkin, Clark, et~al.]{clip}
Alec Radford, Jong~Wook Kim, Chris Hallacy, Aditya Ramesh, Gabriel Goh, Sandhini Agarwal, Girish Sastry, Amanda Askell, Pamela Mishkin, Jack Clark, et~al.
\newblock Learning transferable visual models from natural language supervision.
\newblock In \emph{Proc. International Conference on Machine Learning (ICML)}, 2021.

\bibitem[Raffel et~al.(2020)Raffel, Shazeer, Roberts, Lee, Narang, Matena, Zhou, Li, and Liu]{2020t5}
Colin Raffel, Noam Shazeer, Adam Roberts, Katherine Lee, Sharan Narang, Michael Matena, Yanqi Zhou, Wei Li, and Peter~J. Liu.
\newblock Exploring the limits of transfer learning with a unified text-to-text transformer.
\newblock \emph{Journal of Machine Learning Research (JMLR)}, 2020.

\bibitem[Rajput et~al.(2023)Rajput, Mehta, Singh, Hulikal~Keshavan, Vu, Heldt, Hong, Tay, Tran, Samost, et~al.]{rajput2023recommender}
Shashank Rajput, Nikhil Mehta, Anima Singh, Raghunandan Hulikal~Keshavan, Trung Vu, Lukasz Heldt, Lichan Hong, Yi Tay, Vinh Tran, Jonah Samost, et~al.
\newblock Recommender systems with generative retrieval.
\newblock \emph{Proc. Neural Information Processing Systems (NeurIPS)}, 2023.

\bibitem[Razavi et~al.(2019)Razavi, Van~den Oord, and Vinyals]{razavi2019generating}
Ali Razavi, Aaron Van~den Oord, and Oriol Vinyals.
\newblock Generating diverse high-fidelity images with vq-vae-2.
\newblock In \emph{Proc. Neural Information Processing Systems (NeurIPS)}, 2019.

\bibitem[Saito et~al.(2023)Saito, Sohn, Zhang, Li, Lee, Saenko, and Pfister]{saito2023pic2word}
Kuniaki Saito, Kihyuk Sohn, Xiang Zhang, Chun-Liang Li, Chen-Yu Lee, Kate Saenko, and Tomas Pfister.
\newblock Pic2word: Mapping pictures to words for zero-shot composed image retrieval.
\newblock In \emph{Proc. IEEE Conference on Computer Vision and Pattern Recognition (CVPR)}, 2023.

\bibitem[Singhal et~al.(2001)]{singhal2001modern}
Amit Singhal et~al.
\newblock Modern information retrieval: A brief overview.
\newblock \emph{IEEE Data Eng. Bull.}, 2001.

\bibitem[Sohn(2016)]{Sohn_nips2016}
Kihyuk Sohn.
\newblock Improved deep metric learning with multi-class n-pair loss objective.
\newblock In \emph{Proc. Neural Information Processing Systems (NeurIPS)}, 2016.

\bibitem[Song et~al.(2016)Song, Xiang, Jegelka, and Savarese]{songCVPR16}
Hyun~Oh Song, Yu Xiang, Stefanie Jegelka, and Silvio Savarese.
\newblock Deep metric learning via lifted structured feature embedding.
\newblock In \emph{Proc. IEEE Conference on Computer Vision and Pattern Recognition (CVPR)}, 2016.

\bibitem[Sun et~al.(2024)Sun, Yan, Chen, Wang, Zhu, Ren, Chen, Yin, Rijke, and Ren]{sun2024learning}
Weiwei Sun, Lingyong Yan, Zheng Chen, Shuaiqiang Wang, Haichao Zhu, Pengjie Ren, Zhumin Chen, Dawei Yin, Maarten Rijke, and Zhaochun Ren.
\newblock Learning to tokenize for generative retrieval.
\newblock \emph{Proc. Neural Information Processing Systems (NeurIPS)}, 2024.

\bibitem[Tang et~al.(2023)Tang, Zhang, Guo, Chen, Zhu, Wang, Yin, and Cheng]{tang2023semantic}
Yubao Tang, Ruqing Zhang, Jiafeng Guo, Jiangui Chen, Zuowei Zhu, Shuaiqiang Wang, Dawei Yin, and Xueqi Cheng.
\newblock Semantic-enhanced differentiable search index inspired by learning strategies.
\newblock In \emph{Proceedings of the 29th ACM SIGKDD Conference on Knowledge Discovery and Data Mining}, 2023.

\bibitem[Tay et~al.(2022)Tay, Tran, Dehghani, Ni, Bahri, Mehta, Qin, Hui, Zhao, Gupta, et~al.]{tay2022transformer}
Yi Tay, Vinh Tran, Mostafa Dehghani, Jianmo Ni, Dara Bahri, Harsh Mehta, Zhen Qin, Kai Hui, Zhe Zhao, Jai Gupta, et~al.
\newblock Transformer memory as a differentiable search index.
\newblock In \emph{Proc. Neural Information Processing Systems (NeurIPS)}, 2022.

\bibitem[Wang et~al.(2019)Wang, Han, Huang, Dong, and Scott]{wang2019multi}
Xun Wang, Xintong Han, Weilin Huang, Dengke Dong, and Matthew~R Scott.
\newblock Multi-similarity loss with general pair weighting for deep metric learning.
\newblock In \emph{Proc. IEEE Conference on Computer Vision and Pattern Recognition (CVPR)}, 2019.

\bibitem[Wang et~al.(2022{\natexlab{a}})Wang, Hou, Wang, Miao, Wu, Chen, Xia, Chi, Zhao, Liu, et~al.]{wang2022nci}
Yujing Wang, Yingyan Hou, Haonan Wang, Ziming Miao, Shibin Wu, Qi Chen, Yuqing Xia, Chengmin Chi, Guoshuai Zhao, Zheng Liu, et~al.
\newblock A neural corpus indexer for document retrieval.
\newblock \emph{Proc. Neural Information Processing Systems (NeurIPS)}, 2022{\natexlab{a}}.

\bibitem[Wang et~al.(2022{\natexlab{b}})Wang, Hou, Wang, Miao, Wu, Chen, Xia, Chi, Zhao, Liu, et~al.]{wang2022neural}
Yujing Wang, Yingyan Hou, Haonan Wang, Ziming Miao, Shibin Wu, Qi Chen, Yuqing Xia, Chengmin Chi, Guoshuai Zhao, Zheng Liu, et~al.
\newblock A neural corpus indexer for document retrieval.
\newblock \emph{Proc. Neural Information Processing Systems (NeurIPS)}, 2022{\natexlab{b}}.

\bibitem[Wei et~al.(2024)Wei, Chen, Chen, Hu, Zhang, Fu, Ritter, and Chen]{uniir}
Cong Wei, Yang Chen, Haonan Chen, Hexiang Hu, Ge Zhang, Jie Fu, Alan Ritter, and Wenhu Chen.
\newblock Uniir: Training and benchmarking universal multimodal information retrievers.
\newblock In \emph{Proc. European Conference on Computer Vision (ECCV)}, 2024.

\bibitem[Wei et~al.(2020)Wei, Xu, Yang, Ji, Wang, and Shen]{wei2020universal}
Jiwei Wei, Xing Xu, Yang Yang, Yanli Ji, Zheng Wang, and Heng~Tao Shen.
\newblock Universal weighting metric learning for cross-modal matching.
\newblock In \emph{Proc. IEEE Conference on Computer Vision and Pattern Recognition (CVPR)}, 2020.

\bibitem[Wu et~al.(2021)Wu, Gao, Guo, Al-Halah, Rennie, Grauman, and Feris]{wu2021fashion}
Hui Wu, Yupeng Gao, Xiaoxiao Guo, Ziad Al-Halah, Steven Rennie, Kristen Grauman, and Rogerio Feris.
\newblock Fashion iq: A new dataset towards retrieving images by natural language feedback.
\newblock In \emph{Proc. IEEE Conference on Computer Vision and Pattern Recognition (CVPR)}, 2021.

\bibitem[Young et~al.(2014)Young, Lai, Hodosh, and Hockenmaier]{Flickr30k_b}
Peter Young, Alice Lai, Micah Hodosh, and Julia Hockenmaier.
\newblock From image descriptions to visual denotations: New similarity metrics for semantic inference over event descriptions.
\newblock \emph{Transactions of the Association for Computational Linguistics}, 2014.

\bibitem[Yu and Tao(2019)]{Yu_2019_ICCV}
Baosheng Yu and Dacheng Tao.
\newblock Deep metric learning with tuplet margin loss.
\newblock In \emph{Proc. IEEE International Conference on Computer Vision (ICCV)}, 2019.

\bibitem[Zeghidour et~al.(2021)Zeghidour, Luebs, Omran, Skoglund, and Tagliasacchi]{zeghidour2021soundstream}
Neil Zeghidour, Alejandro Luebs, Ahmed Omran, Jan Skoglund, and Marco Tagliasacchi.
\newblock Soundstream: An end-to-end neural audio codec.
\newblock \emph{IEEE/ACM Transactions on Audio, Speech, and Language Processing}, 2021.

\bibitem[Zhai et~al.(2023)Zhai, Mustafa, Kolesnikov, and Beyer]{zhai2023siglip}
Xiaohua Zhai, Basil Mustafa, Alexander Kolesnikov, and Lucas Beyer.
\newblock Sigmoid loss for language image pre-training.
\newblock In \emph{Proc. IEEE International Conference on Computer Vision (ICCV)}, 2023.

\bibitem[Zhang et~al.(2024)Zhang, Zhang, Chen, Wang, Chen, Xie, Sun, Deng, Zhang, Yang, et~al.]{zhang2023irgen}
Yidan Zhang, Ting Zhang, Dong Chen, Yujing Wang, Qi Chen, Xing Xie, Hao Sun, Weiwei Deng, Qi Zhang, Fan Yang, et~al.
\newblock Irgen: Generative modeling for image retrieval.
\newblock In \emph{Proc. European Conference on Computer Vision (ECCV)}, 2024.

\bibitem[Zhu et~al.(2024)Zhu, Huang, Ding, Yang, Chen, Zhou, Neiman, Xie, Tran, Yao, et~al.]{zhu2024bringing}
Xinliang Zhu, Sheng-Wei Huang, Han Ding, Jinyu Yang, Kelvin Chen, Tao Zhou, Tal Neiman, Ouye Xie, Son Tran, Benjamin Yao, et~al.
\newblock Bringing multimodality to amazon visual search system.
\newblock In \emph{Proceedings of the 30th ACM SIGKDD Conference on Knowledge Discovery and Data Mining}, 2024.

\end{thebibliography}
}

\clearpage
\renewcommand\thesection{\Alph{section}}
\setcounter{section}{0}
\appendix

\section*{Appendix} In this appendix, we present additional experimental results and detailed analyses that could not be included in the main paper due to space limitations. Section~\ref{sup:mbeir} provides an overview of the M-BEIR dataset. Section~\ref{sup:further_analysis} delves into storage and training efficiency. Section~\ref{sup:additional_experiments} offers ablation studies on contrastive loss, modality encoding, beam search, and decoder size. Section~\ref{sup:additional_quantitative} presents further experiments on codebook configurations alongside quantitative evaluations across multiple benchmarks. Finally, Section~\ref{sup:additional_visualization} shows additional visualizations of our modality-decoupled semantic quantization process, demonstrating its capability to capture semantic details in a coarse-to-fine manner.

\begin{table*}[t!]
\centering
\fontsize{7.5}{9}\selectfont
\scalebox{1.0}{
\begin{tabularx}{\textwidth}{ 
   p{0.14\textwidth} 
   >{\centering\arraybackslash}p{0.1\textwidth} 
   >{\centering\arraybackslash}p{0.1\textwidth} 
   >{\centering\arraybackslash}X 
   >{\centering\arraybackslash}X 
   >{\centering\arraybackslash}X 
   >{\centering\arraybackslash}X
   >{\centering\arraybackslash}X
   >{\centering\arraybackslash}X
   >{\centering\arraybackslash}p{0.08\textwidth}
}
\toprule
\multicolumn{1}{l}{\textbf{Task}} & \textbf{Dataset} & \textbf{Domain} & \multicolumn{3}{c}{\textbf{\# Query}} & \multicolumn{3}{c}{\textbf{\# Rel./Query}} & \textbf{\# Candid.} \\ [-0.2ex]
 \cmidrule(lr){4-6} \cmidrule(lr){7-9}
(query $\to$ candidate) & & & Train & Val & Test & Train & Val & Test & \\ [-0.2ex]
\midrule
\multirow{3}{*}[0mm]{1. $q_t \to c_i$} & VisualNews~\cite{liu2017visual} & News & 99K & 20K & 20K & 1.0 & 1.0 & 1.0 & 542K \\
& MSCOCO~\cite{Mscoco} & Misc. & 100K & 24.8K & 24.8K & 1.0 & 1.0 & 1.0 & 5K \\
 & Fashion200K~\cite{han2017automatic} & Fashion & 15K & 1.7K & 1.7K & 3.3 & 3.1 & 2.8 & 201K \\
\midrule
 2. $q_t \to c_t$ & WebQA~\cite{chang2022webqa} & Wiki & 16K & 1.7K & 2.4K & 2.0 & 2.0 & 2.0 & 544K \\
\midrule
 \multirow{2}{*}[0mm]{3. $q_t \to (c_i, c_t)$} & EDIS~\cite{liu2023edis} & News & 26K & 3.2K & 3.2K & 2.6 & 2.6 & 2.6 & 1M \\
 & WebQA~\cite{chang2022webqa} & Wiki & 17K & 1.7K & 2.5K & 1.4 & 1.4 & 1.4 & 403K \\
\midrule
 \multirow{3}{*}[0mm]{4. $q_i \to c_t$} & VisualNews~\cite{liu2017visual} & News & 100K & 20K & 20K & 1.0 & 1.0 & 1.0 & 537K \\
& MSCOCO~\cite{Mscoco} & Misc. & 113K & 5K & 5K & 5.0 & 5.0 & 5.0 & 25K \\
 & Fashion200K~\cite{han2017automatic} & Fashion & 15K & 4.8K & 4.8K & 1.0 & 1.0 & 1.0 & 61K \\
\midrule
 5. $q_i \to c_i$ & NIGHTS~\cite{fu2023dreamsim} & Misc. & 16K & 2K & 2K & 1.0 & 1.0 & 1.0 & 40K \\
\midrule
 \multirow{2}{*}[0mm]{6. $(q_i, q_t) \to c_t$} & OVEN~\cite{hu2023open} & Wiki & 150K & 50K & 50K & 8.5 & 10.0 & 9.9 & 676K \\
 & InfoSeek~\cite{chen2023can} & Wiki & 141K & 11K & 11K & 6.8 & 6.7 & 6.5 & 611K \\
\midrule
\multirow{2}{*}[0mm]{7. $(q_i, q_t) \to c_i$} & FashionIQ~\cite{wu2021fashion} & Fashion & 16K & 2K & 6K & 1.0 & 1.0 & 1.0 & 74K \\
& CIRR~\cite{liu2021image} & Misc. & 26K & 2K & 4K & 1.0 & 1.0 & 1.0 & 21K \\
\midrule
 \multirow{2}{*}[0mm]{8. $(q_i, q_t) \to (c_i, c_t)$} & OVEN~\cite{hu2023open} & Wiki & 157K & 14.7K & 14.7K & 17.8 & 17.5 & 17.7 & 335K \\
& InfoSeek~\cite{chen2023can} & Wiki & 143K & 17.6K & 17.6K & 9.1 & 7.5 & 7.5 & 481K \\ \midrule
\multicolumn{2}{c}{M-BEIR~\cite{uniir}} & 4 domains & 1.1M & 182K & 190K & 6.5 & 5.9 & 5.7 & 5.6M \\
\bottomrule
\end{tabularx}}
\vspace{-2mm}
\caption{Summary of statistics of M-BEIR. Each row describes a task-specific retrieval setup, including the dataset, domain, the number of queries across Train/Validation/Test splits (\# Query), the average number of relevant labels per query (\# Rel./Query), and the total number of candidates (\# Candid.).
}
\label{tab:dataset_stats}
\end{table*}

\section{Details of M-BEIR Dataset}
\label{sup:mbeir}
The M-BEIR dataset~\cite{uniir} combines 10 datasets to support multimodal retrieval tasks, covering diverse domains such as image-caption retrieval, product search, news, and complex multimodal queries. As summarized in Table~\ref{tab:dataset_stats}, it encompasses a total of 5.6M candidates. It supports eight distinct retrieval tasks, including retrieving images from text, text from images, and matching multimodal queries with corresponding multimodal responses. The dataset spans queries with varying levels of complexity, covering multiple domains such as fashion, news, and general-purpose data.

Each query instance consists of a query \( q \), a set of related positive candidates \( c^+ \), and unrelated negative candidates \( c^- \). To clarify the user's intention, each query is paired with an additional intent description. All queries include at least one positive candidate while including negative candidates is optional.

\noindent{\textbf{VisualNews.}} The VisualNews dataset \cite{liu2017visual} was curated by randomly sampling 200K, 40K, and 40K image-caption pairs for training, validation, and testing, respectively. Tasks include retrieving captions \(\bigs( q_i \to c_t \bigs)\) for a given image and retrieving images \(\bigs( q_t \to c_i \bigs)\) for a given caption. The initial number of candidates of 2.5M entries was reduced to 1M in the M-BEIR dataset, consisting of 500K text and 500K image candidates.

\noindent{\textbf{Fashion200K.}} The Fashion200K dataset \cite{han2017automatic}, comprising 200K images and 60K descriptions, was curated by selecting 30K image-description pairs for training. Tasks include retrieving product descriptions \(\bigs( q_i \to c_t \bigs)\) for a given image and retrieving images \(\bigs( q_t \to c_i \bigs)\) for a given product description. The number of candidates is 260K.

\noindent{\textbf{COCO.}} Using the Karpathy split \cite{Karpathy_mscoco_split}, MS-COCO~\cite{Mscoco} data was converted to support tasks such as retrieving captions \(\bigs( q_i \to c_t \bigs)\) from images and retrieving images \(\bigs( q_t \to c_i \bigs)\) from captions. The dataset includes 113K training instances for image-to-caption retrieval, which was trimmed to 100K in the M-BEIR dataset for efficiency. The number of candidates for testing includes 25K text entries and 5K images, the same as the original test set of COCO.

\noindent{\textbf{WebQA.}} The WebQA dataset \cite{chang2022webqa} links textual questions to images and their corresponding textual answers. Tasks include retrieving answers \(\bigs( q_i \to c_t \bigs)\) based on questions and matching queries \(\bigs( q_i \to (c_i, c_t) \bigs)\) with both images and textual explanations. The number of candidates comprises 400K image-text pairs and 540K text-only candidates.

\noindent{\textbf{EDIS.}} The EDIS dataset \cite{liu2023edis} connects captions to image-headline pairs. Tasks involve matching textual queries \(\bigs( q_i \to (c_i, c_t) \bigs)\) with multimodal pairs consisting of images and their associated text. The number of candidates includes 1M image-headline pairs, and the training set consists of 26K instances.

\noindent{\textbf{NIGHTS.}} The NIGHTS dataset \cite{fu2023dreamsim} pairs reference images with target images. The task focuses on retrieving images \(\bigs( q_i \to c_i \bigs)\) based on a reference image. The dataset contains 16K, 2K, and 2K instances for training, validation, and testing, with a number of candidates of 40K images.

\noindent{\textbf{FashionIQ.}} FashionIQ \cite{wu2021fashion} connects reference images and their textual descriptions to target images. Tasks include retrieving target images \(\bigs( q_i \to c_i \bigs)\) based on reference images and associated descriptions. The dataset includes all images as the number of candidates, with 1.7K instances reserved for validation.

\noindent{\textbf{CIRR.}} CIRR \cite{liu2021image} matches reference images and textual modifications to target images. The task involves retrieving target images \(\bigs( (q_i, q_t) \to c_i \bigs)\) that align with both the reference image and the specified textual modification. The number of candidates comprises all images, with validation and test sets derived from the dataset splits.

\noindent{\textbf{OVEN.}} The OVEN dataset \cite{hu2023open} pairs images with text questions and their corresponding multimodal answers. Tasks include retrieving textual descriptions \(\bigs( (q_i, q_t) \to c_t \bigs)\) for a given query and matching multimodal responses \(\bigs( (q_i, q_t) \to (c_i, c_t) \bigs)\). The dataset originally contained 6M candidates, which were reduced to a 1M number of candidates in the M-BEIR dataset, and training data was trimmed to 120K instances.

\noindent{\textbf{InfoSeek.}} InfoSeek \cite{chen2023can} uses queries consisting of images and related questions paired with textual answers segmented into snippets. Tasks include retrieving text snippets \(\bigs( (q_i, q_t) \to c_t \bigs)\) and matching multimodal pairs \(\bigs( (q_i, q_t) \to (c_i, c_t) \bigs)\) with relevant queries. The processed dataset includes 140K instances each for text and multimodal retrieval tasks, with the number of candidates reduced to 1M in the M-BEIR dataset.

\section{Further Analysis}  
\label{sup:further_analysis}
\subsection{Storage Efficiency Comparison}  

Efficient storage utilization is crucial for large-scale retrieval systems. Table~\ref{tab:storage_efficiency} compares the per-data storage requirements of CLIP and GENIUS, highlighting the significant advantage of quantized representations.  

CLIP, which operates on a 768-dimensional floating-point embedding, requires approximately 3\,KB per data point when stored in 32-bit precision. This can lead to substantial storage costs, particularly in large-scale retrieval scenarios.  
In contrast, GENIUS leverages a compact quantization scheme, encoding each data point using a 2-bit code (for modality separation) and eight 12-bit codes selected from a $2^{12}$-sized codebook. This results in a total storage requirement of only \(2 + (8 \times 12) = 98\) bits, equivalent to 12.25 bytes per data point, which is over a 99\% reduction compared to CLIP.  
For example, indexing one million data points would require around 3\,GB with CLIP, whereas GENIUS would require only 12\,MB. This drastic reduction in storage overhead makes GENIUS highly scalable and cost-efficient for deployment in real-world retrieval applications, especially those handling billions of data points.  

\subsection{Training Efficiency}  

GENIUS offers high training efficiency. When training on 1.1 million samples using 4$\times$RTX3090 GPUs, the CLIP encoder requires 91 hours. In comparison, GENIUS introduces an additional 0.4 hours for quantization and 2 hours for decoder training. As a result, on a per-sample basis, GENIUS is approximately 2.8 times more efficient than GRACE, which, according to reports, trains on 0.1 million samples in 24 hours for the MS-COCO dataset.

\begin{table*}[t!]
\centering
\fontsize{8.2}{12}\selectfont
\begin{tabularx}{\textwidth}
   {
   p{0.07\textwidth} |
    >{\centering\arraybackslash}p{0.5\textwidth} |
   >{\centering\arraybackslash}X}
\hline
\textbf{Model} & \textbf{Representation Format} & \textbf{Storage Cost per Data} \\
\hline
CLIP~\cite{clip}  
& 768-dim floating-point vector (32-bit)  
& $768 \times 32 = 24{,}576$ bits = $3{,}072$ bytes $\approx 3$ KB \\
GENIUS  
& Quantized codes: 1 modality code (2-bit) + 8 semantic codes (12-bit each)  
& $2 + (8 \times 12) = 98$ bits $\approx 12.25$ bytes ($\sim 0.012$ KB) \\
\hline
\end{tabularx}
\caption{Comparison of storage efficiency between CLIP and GENIUS. GENIUS achieves a more than 99\% reduction in storage requirements, significantly enhancing scalability for large-scale retrieval tasks.}
\label{tab:storage_efficiency}
\end{table*}

\section{Additional Experiments}
\label{sup:additional_experiments}
\subsection{Impact of Contrastive Loss in Qunatization}
As shown in Table~\ref{tab:main_compnent} of the main paper, $\mathcal{L}_{\text{cl}}$ plays a crucial role, and its removal from the training of quantitation~(Eq.~\ref{eq:rq_total_loss}) leads to near-zero performance. To analyze how contrastive learning affects the embedding space, we conduct a UMAP visualization~\cite{mcinnes2018umap} of the quantized feature $\hat{z}$ before and after applying contrastive learning $\mathcal{L}_{\text{cl}}$ (Eq.~\ref{eq:loss_cl}). Note that the quantized feature $\hat{z}$ is the reconstructed feature using code embeddings derived from discrete IDs.

\begin{figure} [!t]
\centering
\includegraphics[width = 1.0\columnwidth]{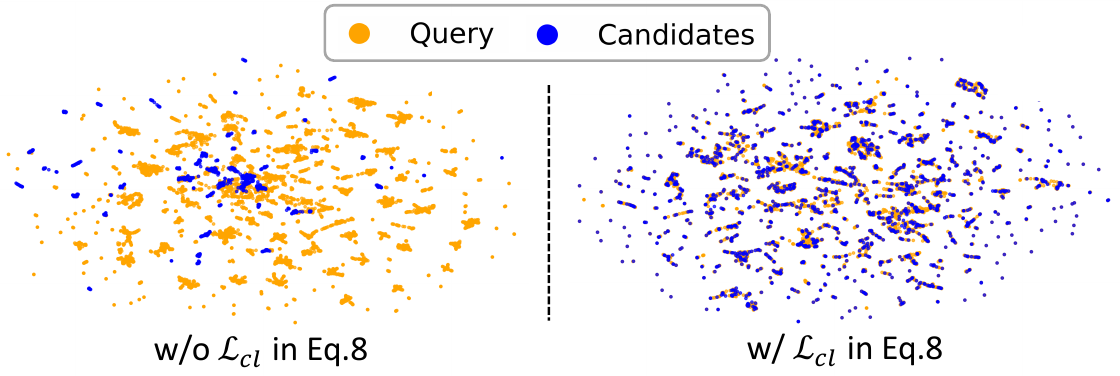}
\vspace{-4mm}
\caption{UMAP visualization of the quantized feature  $\hat{\mathbf{z}}$ before and after contrastive learning $\mathcal{L}_{\text{cl}}$ of Eq.~\ref{eq:loss_cl}} 
\label{fig:umap_contra}
\end{figure}

Fig.~\ref{fig:umap_contra} illustrates that even though residual quantization loss (Eq.~\ref{eq:loss_ml}) is applied, removing contrastive learning results in misalignment between query and target features and causes target features to collapse. This degradation in representation leads to discrete IDs that fail to capture the relations between queries and targets effectively, making it difficult for the decoder to learn it. Furthermore, an excessive number of targets become mapped to a single ID, rendering the retrieval process ineffective and generating semantically inconsistent IDs.
In contrast, when contrastive loss is applied in Eq.~\ref{eq:rq_total_loss}, query-target alignment is preserved despite quantization. This ensures that the semantic information is well-represented within the discrete IDs. As a result, when training the decoder to map queries to targets, it can effectively capture the underlying relations, allowing it to generate meaningful discrete target IDs from queries.

\subsection{Impact of Modality Encoding}
We analyze the impact of modality encoding by comparing different quantization strategies in Table~\ref{tab:ablation_quantization}: modality-decoupled quantization, classifier-based modality encoding, and residual quantization without a modality code.

Modality-decoupled quantization achieves the best performance among the three approaches. While classifier-based encoding successfully differentiates modalities, it does not integrate modality information within the quantization process. As a result, modality and semantic information are mixed within the discrete codes, limiting their representational capacity. In contrast, modality-decoupled quantization explicitly separates modality information by assigning the first code to modality while using the remaining codes for semantics, leading to a more structured and expressive representation.

The baseline without modality encoding, which does not explicitly separate modalities, further demonstrates that failing to encode modality weakens retrieval performance. These findings emphasize that modality-decoupled quantization provides a unified approach for handling multiple modalities in generative retrieval, offering a more effective discrete ID representation.


\subsection{Impact of Beam Search}
We conduct an ablation study to examine the impact of beam size on retrieval performance and efficiency across various tasks. As shown in Table~\ref{suptab:beam}, increasing the beam size significantly improves Recall@5. For instance, on the COCO dataset for text-to-image retrieval, Recall@5 increases from 24.2\% at a beam size of 1 to 68.2\% at a beam size of 50. Similar trends are observed for image-to-text retrieval on COCO and image-to-image retrieval on CIRR. The improvement is even more pronounced on the WebQA dataset, which contains knowledge-intensive data in Wikipedia based on long sentence queries. Recall@5 for text-to-text retrieval increases from 5.1\% at a beam size of 1 to 32.8\% at a beam size of 50. This substantial gain is attributed to the expanded search space provided by larger beam sizes, allowing the model to handle better the complexity and richness of the queries in WebQA.

However, larger beam sizes increase the computational load, resulting in higher latency. Based on our measurements of the text-to-image retrieval task, retrieval speed decreases from 19.6 queries per second at a beam size of 30 to 11.9 queries per second at a beam size of 50. This trade-off between performance and efficiency is a fundamental consideration when deploying generative models using beam search. Selecting an appropriate beam size requires balancing the need for higher recall against the constraints of computational resources and application-specific latency requirements.

\begin{table}[t!]
\centering
\fontsize{7.5}{9}\selectfont
\setlength{\tabcolsep}{2pt} 
\scalebox{0.95}{
\begin{tabularx}{0.499\textwidth}{ 
   p{0.22\textwidth} |
   >{\centering\arraybackslash}X |
   >{\centering\arraybackslash}X |
   >{\centering\arraybackslash}X |
   >{\centering\arraybackslash}X |
   >{\centering\arraybackslash}X
   } 
\toprule
 & \multicolumn{2}{c}{\textbf{COCO}} & \multicolumn{2}{c}{\textbf{WebQA}} & \textbf{CIRR} \\ 
\cmidrule(lr){2-3} \cmidrule(lr){4-5} \cmidrule(lr){6-6} 
\textbf{Method} & \fontsize{7}{8}\selectfont{T2I} & \fontsize{7}{8}\selectfont{I2T} & \fontsize{7}{8}\selectfont{T2T} & \fontsize{7}{8}\selectfont{T2(I,T)} & \fontsize{7}{8}\selectfont{(I,T)2I} 
 \\  \midrule
Modality-decoupled quantization & \textbf{55.4} & \textbf{82.7}& \textbf{28.3}&\textbf{47.1}&\textbf{20.5} \\
Classifier-based modality encoding & 48.9 & 79.2 & 25.7 & 37.5 & 20.3 \\
{RQ w/o modality-code}& 20.2 & 73.2 & 25.9 & 34.3 & 18.3 \\
\bottomrule
\end{tabularx}
}
\caption{Ablation study on modality encoding approach (universal retrieval, R@5).}
\label{tab:ablation_quantization}
\end{table}

\subsection{Impact of Decoder Size}
We analyze the effect of the decoder size on retrieval performance. Table~\ref{suptab:decoder} presents the results using T5 decoders~\cite{2020t5} of varying sizes: T5-small (30M parameters), T5-base (110M parameters), and T5-large (400M parameters). Increasing the decoder size generally enhances performance on tasks like COCO and WebQA. On COCO text-to-image retrieval, Recall@5 improves from 65.3\% with T5-small to 67.9\% with T5-base. On WebQA, performance increases consistently with decoder size, reaching 32.4\% Recall@5 with T5-large, which is beneficial for handling complex sentences in WebQA.
However, on the CIRR dataset, which involves complex relational reasoning in image-to-image retrieval, performance declines slightly with T5-base and drops sharply to 7.1\% with T5-large. This suggests that larger models may overfit or struggle with optimization on certain tasks, especially those that do not benefit from increased model capacity. Therefore, we adopt T5-small as the default decoder for its effective trade-off between retrieval performance and computational efficiency.

\begin{table}[t!]
\centering
\fontsize{8}{10}\selectfont
\scalebox{0.95}{
\begin{tabularx}{0.499\textwidth}{ 
   >{\centering\arraybackslash}p{0.077\textwidth} |
   >{\centering\arraybackslash}X |
   >{\centering\arraybackslash}X |
   >{\centering\arraybackslash}X |
   >{\centering\arraybackslash}X |
   >{\centering\arraybackslash}X
   } 
\toprule
 & \multicolumn{2}{c}{\textbf{COCO}} & \multicolumn{2}{c}{\textbf{WebQA}} & \textbf{CIRR} \\ [-0.4ex] 
\cmidrule(lr){2-3} \cmidrule(lr){4-5} \cmidrule(lr){6-6} 
\textbf{Beam Size} & \fontsize{7}{8}\selectfont{T2I} & \fontsize{7}{8}\selectfont{I2T} & \fontsize{7}{8}\selectfont{T2T} & \fontsize{7}{8}\selectfont{T2(I,T)} & \fontsize{7}{8}\selectfont{(I,T)2I} 
 \\ [-0.4ex]  \midrule
1 & 24.2& 41.6& 5.1&10.4 &4.9 \\
5 & 55.6& 79.1& 15.9& 32.3& 18.0\\
10 & 62.8& 82.8& 22.4& 40.0&20.4 \\
20 & {66.5}& \textbf{83.7}&28.3 &45.1 &\textbf{21.1} \\ 
\ccol 30 & \ccol {65.3} & \ccol{83.4} & \ccol {28.8} & \ccol{47.4}&\ccol 21.0 \\ 
50 &\textbf{68.2} &83.3 & \textbf{32.8}& \textbf{50.0}& 21.0\\ 
\bottomrule
\end{tabularx}
}
\vspace{-1mm}
\caption{Ablation over beam size (task-specific information retrieval, R@5). The default setting of our method is highlighted in \colorbox{grey}{{grey box}}.}
\label{suptab:beam}
\vspace{-2mm}
\end{table}

\begin{table}[t!]
\centering
\fontsize{8}{10}\selectfont
\scalebox{0.95}{
\begin{tabularx}
    {0.499\textwidth}{ 
   >{\centering\arraybackslash}p{0.074\textwidth} |
   >{\centering\arraybackslash}p{0.07\textwidth} |
   >{\centering\arraybackslash}X |
   >{\centering\arraybackslash}X |
   >{\centering\arraybackslash}X |
   >{\centering\arraybackslash}X |
   >{\centering\arraybackslash}X
   } 
\toprule
 & & \multicolumn{2}{c}{\textbf{COCO}} & \multicolumn{2}{c}{\textbf{WebQA}} & \textbf{CIRR} \\ [-0.4ex] 
\cmidrule(lr){3-4} \cmidrule(lr){5-6} \cmidrule(lr){7-7} 
\textbf{Decoder} & \textbf{\# Params} & \fontsize{7}{8}\selectfont{T2I} & \fontsize{7}{8}\selectfont{I2T} & \fontsize{7}{8}\selectfont{T2T} & \fontsize{7}{8}\selectfont{T2(I,T)} & \fontsize{7}{8}\selectfont{(I,T)2I} 
 \\ [-0.4ex]  \midrule
\ccol T5-small & \ccol 30M & \ccol {65.3} & \ccol{83.4} & \ccol {28.8} & \ccol{47.4}&\ccol \textbf{21.0} \\ 
 T5-base &  110M & \textbf{67.9} &\textbf{83.5} &  31.6 &48.0 &18.3  \\ 
 T5-large & 400M &   {67.2} & {83.2} &  \textbf{{32.4}} & \textbf{{50.4}}& 7.1 \\ 
\bottomrule
\end{tabularx}
}
\vspace{-1mm}
\caption{Ablation over decoder size (task-specific information retrieval, R@5). The default setting of our method is highlighted in \colorbox{grey}{{grey box}}.}
\label{suptab:decoder}
\vspace{-3mm}
\end{table}

\begin{table}[t!]
\centering
\fontsize{8}{10}\selectfont
\scalebox{0.95}{
\begin{tabularx}{0.499\textwidth}{ 
   p{0.12\textwidth} |
   >{\centering\arraybackslash}X |
   >{\centering\arraybackslash}X |
   >{\centering\arraybackslash}X |
   >{\centering\arraybackslash}X |
   >{\centering\arraybackslash}X
   } 
\toprule
 & \multicolumn{2}{c}{\textbf{COCO}} & \multicolumn{2}{c}{\textbf{WebQA}} & \textbf{CIRR} \\ [-0.4ex] 
\cmidrule(lr){2-3} \cmidrule(lr){4-5} \cmidrule(lr){6-6} 
\textbf{$K \times M$} & \fontsize{7}{8}\selectfont{T2I} & \fontsize{7}{8}\selectfont{I2T} & \fontsize{7}{8}\selectfont{T2T} & \fontsize{7}{8}\selectfont{T2(I,T)} & \fontsize{7}{8}\selectfont{(I,T)2I} 
 \\ [-0.4ex]  \midrule
\ccol 4096$\times$ 9 &\ccol  \textbf{65.3} &\ccol  \textbf{83.4} &\ccol  {28.8}&\ccol  \textbf{47.4}& \ccol 21.0 \\ 

\underline{8192}$\times$ \underline{17} 
& 59.5 & 81.8& \textbf{30.6}& 44.8& \textbf{26.5}\\ 
\underline{4096}$\times$ \underline{9} (Shared) 
& 18.6& 19.3& 0.2& 1.7& 3.3\\ 
\bottomrule
\end{tabularx}
}
\vspace{-1mm}
\caption{Ablation over codebook size $K$ (except for the first level) and code level $M$ (task-specific information retrieval, R@5). The default codebook size and level are underlined. In the shared configuration, codebooks are shared across all levels except the first. The default setting of our method is highlighted in \colorbox{grey}{{grey box}}.  }
\label{suptab:codebook}
\vspace{-3mm}
\end{table}

\subsection{Further Analysis of Codebook Configuration}
We further investigate the impact of codebook configurations, including codebook size ($K$), code levels ($M$) and shared codebook usage across levels in our modality-decoupled semantic quantization. Table~\ref{suptab:codebook}  shows the results for different configurations. Increasing the codebook size and the number of code levels to $K = 8192$, $M = 17$ does not necessarily improve performance. For instance, on COCO text-to-image retrieval, Recall@5 decreases from 65.3\% to 59.5\%. However, on CIRR, this configuration leads to a significant performance improvement, highlighting the varying impact of codebook size depending on task complexity and modality.
Overly large and fine-grained codebook configurations, while occasionally beneficial, increase the complexity of training the decoder model. 

When using a shared codebook, Recall@5 on COCO drops drastically to 18.6\%. Similar declines are observed across other tasks, indicating that level-specific codebooks are crucial for capturing the unique characteristics of different semantics. These findings highlight the importance of carefully configuring the codebook to ensure effective quantization and retrieval performance.

\section{Additional Quantitative Results}
\label{sup:additional_quantitative}
We present performance evaluations for additional settings not covered in the main paper, including variations in beam size and comparisons with a broader range of baselines.

\subsection{Standard Generative Retrieval Benchmark}
\begin{table}[!t]
\centering
\setlength{\tabcolsep}{4pt}
\fontsize{8}{10}\selectfont
\scalebox{0.98}{
\begin{tabularx}{\columnwidth}{
    p{0.39\columnwidth} 
    >{\centering\arraybackslash}p{0.22\columnwidth} 
    >{\centering\arraybackslash}X
    >{\centering\arraybackslash}X 
    >{\centering\arraybackslash}X
}
\toprule
\textbf{Method} & \textbf{Training Data} &\textbf{R@1} & \textbf{R@5} & \textbf{R@10} \\
\midrule
\multicolumn{5}{c}{\textbf{Flickr30K}} \\ \midrule
GRACE~\cite{li2024grace} (Numeric ID) & Flickr30K & 22.5 & 28.9 & 29.4 \\
GRACE~\cite{li2024grace} (String ID)  & Flickr30K & 30.5 & 39.0 & 40.4 \\
GRACE~\cite{li2024grace} (Semantic ID) & Flickr30K & 22.9 & 34.9 & 37.4 \\
GRACE~\cite{li2024grace} (Structured ID) & Flickr30K & 37.4 & 59.5 & 66.2 \\
IRGen~\cite{zhang2023irgen} & Flickr30K & 49.0 & 68.9 & 72.5 \\
\midrule
\textbf{GENIUS} & M-BEIR & \textit{51.5}$^\dagger$ & \textit{74.6}$^\dagger$ & \textit{80.3}$^\dagger$ \\ 
\textbf{GENIUS}$^{\mathcal{R}}$ & M-BEIR & \textit{63.7}$^\dagger$ & \textit{80.4}$^\dagger$ & \textit{83.2}$^\dagger$ \\ 
\textbf{GENIUS} & Flickr30K & 60.6&  84.0& 90.5 \\ 
\textbf{GENIUS}$^{\mathcal{R}}$  & Flickr30K & \textbf{74.1}& \textbf{92.0}& \textbf{94.8}\\ 
\midrule
\multicolumn{5}{c}{\textbf{COCO}} \\ \midrule
GRACE~\cite{li2024grace} (Numeric ID)& COCO  & 0.03 & 0.14 & 0.28 \\
GRACE~\cite{li2024grace} (String ID) & COCO & 0.12 & 0.37 & 0.88 \\
GRACE~\cite{li2024grace} (Semantic ID) & COCO & 13.3 & 30.4 & 35.9 \\
GRACE~\cite{li2024grace} (Structured ID) & COCO & 16.7 & 39.2 & 50.3 \\
IRGen~\cite{zhang2023irgen} & COCO & 29.6 & 50.7 & 56.3 \\
\midrule
\textbf{GENIUS} & M-BEIR & 40.0 & 65.5 & 76.8 \\ 
\textbf{GENIUS}$^{\mathcal{R}}$ &M-BEIR  & 42.6 & 67.3 & 78.9 \\ 
\textbf{GENIUS} & COCO & 41.2 & 67.8  & 77.8  \\ 
\textbf{GENIUS}$^{\mathcal{R}}$ & COCO & \textbf{46.1} & \textbf{74.0}& \textbf{82.7} \\ 
\bottomrule
\end{tabularx}
}
\vspace{-2mm}
\caption{
\textbf{Comparison of generative retrieval methods on text-to-image retrieval benchmarks.} Results are reported as Recall@k (\%). $\dagger$ indicates zero-shot performance, highlighting the ability of the model to generalize without task-specific fine-tuning.
}
\label{suptab:flickr_coco}
\vspace{-3mm}
\end{table}

We evaluate GENIUS against prior generative retrieval methods, including GRACE and IRGen, on standard text-to-image benchmarks such as Flickr30K and COCO, as summarized in Table~\ref{suptab:flickr_coco}. Unlike GRACE and IRGen, which are specifically designed for text-to-image tasks, GENIUS is originally trained on the M-BEIR benchmark in a multi-task setting, supporting diverse retrieval scenarios while also being capable of task-specific training. Note that Flickr30K is not included in the M-BEIR dataset.

\begin{table*}[!t]
\vspace{-2mm}
\setlength{\tabcolsep}{2pt} 
\centering
\scalebox{1.0}{
\fontsize{7.5}{9}\selectfont 
\centering

\begin{tabularx}{\textwidth}{ 
    p{0.15\textwidth} 
    >{\centering\arraybackslash}p{0.113\textwidth} 
    *{4}{>{\centering\arraybackslash}X >{\centering\arraybackslash}X} 
}
\toprule
\textbf{} & \multirow{1}{*}[-2mm]{\textbf{Fine-tuning}}& 
\multicolumn{2}{c}{\textbf{COCO}} & 
\multicolumn{2}{c}{\textbf{VisualNews}} & 
\multicolumn{2}{c}{\textbf{Fashion200K}} & 
\multicolumn{1}{c}{\textbf{Nights}} & 
\multicolumn{1}{c}{\textbf{EDIS}} \\  [-0.4ex] 
\cmidrule(lr){3-4} \cmidrule(lr){5-6} \cmidrule(lr){7-8} \cmidrule(lr){9-9} \cmidrule(lr){10-10}
 &  & 
T to I & I to T & 
T to I & I to T & 
T to I & I to T & 
I to I &
T to (I,T) \\  [-0.4ex] 
\midrule

\multicolumn{10}{c}{\textit{\textbf{Embedding-based Retrieval}}} \\ [-0.4ex] \midrule
 
CLIP-SF~\cite{uniir} & \multirow{2}{*}{Single Task} &  
81.7 & 89.8  & 
\textbf{43.5} & 42.7 & 
10.7 & 12.0  & 
\textbf{33.5} & 
58.8\\ 
  BLIP-FF~\cite{uniir} &  & 
 77.3 &  86.0 & 
20.0   & 22.4& 
17.1  & 15.6 & 
30.4  &  
38.2\\ \midrule

CLIP-SF~\cite{uniir} & \multirow{2}{*}{Unified Instruction} & 
\textbf{81.1} & \textbf{92.3} & 
42.6 & \textbf{43.1} & 
18.0 & 18.3 & 
32.0 &  
\textbf{59.4}   \\ 
BLIP-FF~\cite{uniir}  & & 
67.5 & 89.9& 
 23.4 & 22.8& 
 \textbf{26.1 }& \textbf{28.9} & 
33.0  &  
 50.9 \\ \midrule
 \multicolumn{10}{c}{\textit{\textbf{Generative Retrieval}}} \\ [-0.4ex] \midrule

  GRACE ~\cite{li2024grace} & \multirow{2}{*}{Single Task} & 
39.5 & -- & 
-- & -- & 
-- & -- & 
-- &  
--  \\ 
  IRGen~\cite{zhang2023irgen} &  &
50.7 & -- & 
-- & -- & 
-- & -- & 
-- &  
--  \\ 

\midrule
\ccol \textbf{GENIUS $\,\,\, (\mathcal{B}=30)$} & \ccol   &   
\ccol {65.5} & \ccol {83.4} & 
\ccol {17.5} & \ccol {17.5} & 
\ccol {13.6} & \ccol {17.0} & 
\ccol {8.4} & \ccol {35.6}  \\

\ccol \textbf{GENIUS$^\mathcal{R} (\mathcal{B}=30)$} & \ccol   & 
\ccol {67.3} & \ccol {89.7} & 
\ccol {23.3} & \ccol {24.0} & 
\ccol {15.2} & \ccol \textbf{18.9} & 
\ccol {29.0} & \ccol {41.4}  \\

  \ccol \textbf{GENIUS $\,\,\, (\mathcal{B}=50)$}& \ccol  \multirow{1}{*}[1.5mm]{Unified Instruction} & 
  \ccol  68.1 &  \ccol  83.2 & 
 \ccol  18.5 & \ccol  18.7 & 
 \ccol  13.7 &  \ccol  12.8 & 
 \ccol  8.2 &  
 \ccol  37.0 \\
 
\ccol \textbf{GENIUS$^\mathcal{R} (\mathcal{B}=50)$} & \ccol   & \textbf{78.0}
\ccol & \textbf{91.1}\ccol  & 
\ccol \textbf{27.4} &\textbf{28.4} \ccol  & 
\ccol  \textbf{16.2} & 16.3 \ccol  & 
\ccol \textbf{30.2} & \textbf{44.3} \ccol  \\

\toprule
\textbf{} & {\multirow{1}{*}[-2mm]{\textbf{Fine-tuning}}}& 
\multicolumn{2}{c}{\textbf{WebQA}} & 
\multicolumn{2}{c}{\textbf{OVEN}} & 
\multicolumn{2}{c}{\textbf{InfoSeek}} & 
\multicolumn{1}{c}{\textbf{FashionIQ}} & 
\multicolumn{1}{c}{\textbf{CIRR}} \\  [-0.4ex] 
\cmidrule(lr){3-4} \cmidrule(lr){5-6} \cmidrule(lr){7-8} \cmidrule(lr){9-9} \cmidrule(lr){10-10} &  &
T to T & T to (I,T) & 
(I,T) to T & (I,T) to (I,T) & 
(I,T) to T & (I,T) to (I,T) & 
(I,T) to I &
(I,T) to I \\  [-0.4ex] 
\midrule

 \multicolumn{10}{c}{\textit{\textbf{Embedding-based Retrieval}}} \\ [-0.4ex] \midrule

 CLIP-SF~\cite{uniir} & \multirow{2}{*}{Single Task} & 
81.7 & 76.3  & 
45.4  &{66.2}  & 
23.5 & 47.4 & 
25.9 &  
52.0   \\ 
 BLIP-FF~\cite{uniir} &  &
67.5 & 67.8& 
33.8 & 49.9 & 
18.5 &32.3  & 
3.0  &  
13.9\\ \midrule

CLIP-SF~\cite{uniir} & \multirow{2}{*}{Unified Instruction} &  
\textbf{84.7 }& 78.7 & 
\textbf{45.5} &\textbf{67.6} & 
\textbf{23.9}&\textbf{48.9} & 
24.4&  
44.6 \\ 
 BLIP-FF~\cite{uniir}  && 
 80.0 &\textbf{79.8} & 
 41.0 & 55.8 & 
 22.4 & 33.0 & 
\textbf{29.2} &  
\textbf{52.2} \\ 
\midrule
 \multicolumn{10}{c}{\textit{\textbf{Generative Retrieval}}} \\ [-0.4ex] \midrule

 \ccol \textbf{GENIUS$\,\,\,\, (\mathcal{B}=30)$} & \ccol  & 
 \ccol  {28.8} &  \ccol  {47.4} & 
 \ccol  {34.9} & \ccol   {34.6} & 
 \ccol  {12.4} &  \ccol  {15.1} & 
 \ccol {12.8} &  
 \ccol {21.0}  \\

  \ccol \textbf{GENIUS$^\mathcal{R} (\mathcal{B}=30)$}& \ccol   &  
  \ccol  {36.3} &  \ccol  {54.9} & 
 \ccol  {36.6} & \ccol   {35.0} & 
 \ccol  {18.0} &  \ccol  {26.7} & 
 \ccol {17.5} &  
 \ccol {35.5}  \\

   \ccol \textbf{GENIUS $\,\,\,\, (\mathcal{B}=50)$}& \ccol  \multirow{1}{*}[1.5mm]{Unified Instruction} &  
  \ccol  32.5 &  \ccol  49.7 & 
 \ccol  36.6 & \ccol 36.4 & 
 \ccol 11.2  &  \ccol 14.6  & 
 \ccol  13.2 &  
 \ccol 20.7  \\

\ccol \textbf{GENIUS$^\mathcal{R} (\mathcal{B}=50)$} 
& \ccol 
& \ccol \textbf{44.6}   &\textbf{60.6} \ccol  
& \ccol \textbf{41.9} & \ccol  \textbf{52.5}
& \ccol  \textbf{20.7} & \ccol  \textbf{30.1}
& \ccol \textbf{19.3} & \ccol \textbf{39.5} \\

\bottomrule
\end{tabularx}
}
\vspace{-2mm}
\caption{
\textbf{Task-specific Information Retrieval.} Recall@5 results of single-task and unified instruction fine-tuning methods on the M-BEIR dataset, except Fashion200K and FashionIQ, where Recall@10 is reported. $\mathcal{B}$ represents the beam size, and $\mathcal{R}$ indicates re-ranking based on embedding vectors within the predicted candidate set. I and T denote image and text modalities, respectively, and (I,T) indicates the retrieval direction for image-to-text or text-to-image tasks.
}
\label{suptab:results_local_full}
\end{table*}

On Flickr30K, GENIUS trained with M-BEIR achieves an impressive zero-shot Recall@5 of 74.1\%, surpassing GRACE by over 15 percentage points, despite having never seen the dataset during training. When fine-tuned exclusively on Flickr30K and combined with re-ranking, GENIUS further improves its performance to a Recall@5 of 92.0\%, setting a new state-of-the-art for generative retrieval on this benchmark.
On COCO, GENIUS trained with M-BEIR achieves a Recall@5 of 65.5\%, significantly outperforming GRACE (39.2\%) and IRGen (50.7\%). When trained solely on COCO, GENIUS improves further to a Recall@5 of 74.0\%. 
These results highlight the generalization ability of GENIUS to unseen datasets within a multi-task learning framework. Although M-BEIR includes domains similar to Flickr30K (\eg, COCO), GENIUS achieves zero-shot performance that surpasses models specifically trained on the same domain. Furthermore, GENIUS excels in task-specific scenarios, achieving superior performance when trained on individual datasets and achieving state-of-the-art results.

\subsection{Dataset-Specific Retrieval}
Table~\ref{suptab:results_local_full} summarizes the performance of GENIUS across various retrieval tasks, demonstrating its ability to outperform prior generative methods and achieve results close to state-of-the-art embedding-based baselines in specific tasks.
For text-to-image retrieval on COCO, GENIUS achieves a Recall@5 of 65.5\% with a beam size of 30, significantly surpassing IRGen at 50.7\%. With embedding-based re-ranking, performance improves to 78.0\%, narrowing the gap with CLIP-SF, which achieves 81.7\%. In image-to-text retrieval on COCO, GENIUS achieves a Recall@5 of 91.1\% with re-ranking and a beam size of 50, nearly matching the 92.3\% of CLIP-SF.

\begin{table*}[!t]
\setlength{\tabcolsep}{2pt}
\centering
\fontsize{7.5}{9}\selectfont 
\scalebox{1.0}{ 
\begin{tabularx}{\textwidth}{ 
    p{0.34\columnwidth} 
    p{0.18\columnwidth} 
    >{\centering\arraybackslash}X
    >{\centering\arraybackslash}X
    >{\centering\arraybackslash}X
    >{\centering\arraybackslash}X 
    >{\centering\arraybackslash}X 
    >{\centering\arraybackslash}X
    >{\centering\arraybackslash}X
    >{\centering\arraybackslash}X
}
\toprule
\multicolumn{2}{l}{} & \multicolumn{4}{c}{\textbf{Embedding-based Retrieval}} & \multicolumn{4}{c}{\textbf{Generative Retrieval}}   \\
\cmidrule(lr){3-6} \cmidrule(lr){7-10}
\multirow{1}{*}[1mm]{\textbf{Task}} & \multirow{1}{*}[1mm]{\textbf{Dataset}} & 
\multirow{1}{*}[-1.5mm]{${\text{CLIP}_\text{SF}}$}
 & \multirow{1}{*}[-1.5mm]{${\text{CLIP}_\text{FF}}$}
 & \multirow{1}{*}[-1.5mm]{${\text{BLIP}_\text{SF}}$}
 & \multirow{1}{*}[-1.5mm]{${\text{BLIP}_\text{FF}}$}
 & \textbf{GENIUS $(\mathcal{B}=30)$} & \textbf{GENIUS$^\mathcal{R}$ $(\mathcal{B}=30)$} &
 \textbf{GENIUS $(\mathcal{B}=50)$}  &
\textbf{GENIUS$^\mathcal{R}$  $(\mathcal{B}=50)$} \\
\midrule
\multirow{3}{*}{1. $q_t \to c_i$} 
 & VisualNews & \textbf{42.6}& 28.8 & 20.9 & 23.0 & 18.5& 23.9 & 18.5 & \textbf{27.3}\\
 & MSCOCO & \textbf{77.9} & 74.7 & 71.6 & 75.6 & 55.4 & 64.8 &  55.1 & \textbf{68.0}\\
 & Fashion200K & 17.8 & 15.5 & 24.3 & \textbf{25.4} & 13.6 & 14.7 & 13.7 & \textbf{16.2}\\
\midrule
\multirow{1}{*}{2. $q_t \to c_t$} 
 & WebQA & \textbf{84.7}& 78.4 & 78.9 & 79.5 & 28.3 & 36.5 & 31.1 & \textbf{42.9}\\
\midrule
\multirow{2}{*}{3. $q_t \to (c_i, c_t)$} 
 & EDIS &\textbf{59.4} & 50.0 & 47.2 & 50.3 & 35.4 & 41.4 & 36.6& \textbf{44.1}\\
 & WebQA & 78.8 & 75.3 & 76.8 &\textbf{ 79.7} & 47.1 & 55.8 & 49.0 & \textbf{59.7}\\
\midrule
\multirow{3}{*}{4. $q_i \to c_t$} 
 & VisualNews &\textbf{42.8}& 28.6 & 19.4 & 21.1 & 17.3 & 23.2 & 18.4 & \textbf{26.8}\\
 & MSCOCO & \textbf{92.3}& 89.0 & 88.2 & 88.8 & 82.7 & 89.4 &82.7 & \textbf{90.6}\\
 & Fashion200K & 17.9 & 13.7 & 24.3 & \textbf{27.6} & 12.2 & 14.8 & 12.8 & \textbf{16.2} \\
\midrule
\multirow{1}{*}{5. $q_i \to c_i$} 
 & NIGHTS & 32.0 & 31.9 &\textbf{33.4} & 33.0 & 8.4  & 28.8 & 8.1& \textbf{30.2}\\
\midrule
\multirow{2}{*}{6. $(q_i, q_t) \to c_t$} 
 & OVEN &\textbf{39.2} & 34.7 & 35.2 & 38.7 & 34.4 & 37.1 & 34.6 & \textbf{38.0}\\
 & InfoSeek & \textbf{24.0} & 17.5 & 16.7 & 19.7 & 11.1  & 16.6 & 10.4 & \textbf{18.0}\\
\midrule
\multirow{2}{*}{7. $(q_i, q_t) \to c_i$} 
 & FashionIQ & 24.3 & 20.5 & 26.2 & \textbf{28.5} & 12.8 & 17.4 &  18.9&\textbf{19.2} \\
 & CIRR & 43.9 & 40.9 & 43.0 & \textbf{51.4}& 20.5 & 34.9 &20.1 & \textbf{38.3} \\
\midrule
\multirow{2}{*}{8. $(q_i, q_t) \to (c_i, c_t)$} 
 & OVEN &\textbf{60.2} & 55.8 & 51.8 & 57.8 & 36.9 & 40.9 & 36.5& \textbf{48.6} \\
 & InfoSeek & \textbf{44.6} & 36.8 & 25.4 & 27.7 & 14.3 & 25.7 & 14.2 & \textbf{28.6}\\
\midrule
\textbf{Average} & & \textbf{48.9}& 43.3 & 42.7 & 45.5 & 28.1 &35.4 & 28.8 & \textbf{38.3}\\
\bottomrule
\end{tabularx}
}
\vspace{-2mm}
\caption{\textbf{Universal Information Retrieval.} Recall@5 for various tasks on the M-BEIR dataset, retrieved from a global pool across diverse modalities. $\mathcal{B}$ represents the beam size, and $\mathcal{R}$ indicates re-ranking based on embedding vectors within the predicted candidate set.}
\vspace{-3.5mm}
\label{suptab:results_full}
\end{table*}

For relational reasoning tasks in CIRR, GENIUS achieves a Recall@5 of 35.5\% with a beam size of 30. Increasing the beam size to 50 and incorporating re-ranking raises performance to 39.5\%, demonstrating its strength in addressing relational queries. On WebQA, which features knowledge-intensive and long-form queries, embedding-based re-ranking boosts Recall@5 for text-to-text retrieval from 36.3\% to 44.6\%, effectively leveraging additional search space to handle semantically complex data. GENIUS already shows superior performance compared to prior generative methods with beam search alone. Moreover, by combining larger beam sizes with embedding-based re-ranking, GENIUS often achieves performance levels that are competitive with embedding-based state-of-the-art methods.

\subsection{Universal Retrieval}
The universal retrieval performance of GENIUS demonstrates its ability to handle diverse tasks effectively, as shown in Table~\ref{suptab:results_full}. 
Increasing the beam size alone does not always result in significant performance improvements. However, embedding-based re-ranking plays a crucial role in refining candidate sets and enhancing retrieval performance, often enabling GENIUS to approach state-of-the-art performance.

For image-to-text retrieval on MSCOCO, Recall@5 improves from 82.7\% with beam search alone to 90.6\% with re-ranking at a beam size of 50, narrowing the gap with CLIP-SF (92.3\%). This highlights the strength of re-ranking in prioritizing relevant candidates that may not rank highly within the initial beam output. 
Similarly, on the OVEN dataset for image and text pair-to-text retrieval, Recall@5 increases from 34.4\% to 38.0\% with re-ranking at a larger beam size, effectively closing the gap with CLIP-SF (39.2\%). On NIGHTS, which involves image-to-image retrieval, re-ranking produces a substantial improvement, with Recall@5 jumping from 8.4\% to 30.2\% at the largest beam size.
These results indicate that while GENIUS generates strong candidates through beam search, embedding-based re-ranking is essential to achieve competitive performance, especially at larger beam sizes where the expanded search space requires further refinement to prioritize relevance.

\section{More Visualizations of Quantization}
\label{sup:additional_visualization}
To illustrate how our modality-decoupled semantic quantization operates, we provide further visualizations demonstrating its dual properties of modality separation and coarse-to-fine semantic refinement across subsequent levels. These examples highlight the ability of GENIUS to handle multimodal data through structured code, capturing progressively distinct semantic details.

At the \textbf{first level}, codes represent modality distinctions: 0 for images, 1 for text, and 2 for image-text pairs. This clear separation ensures that the retrieval system processes each modality appropriately, which forms the foundation for multimodal data handling.

\begin{figure*} [!t]
\centering
\includegraphics[width = 1.0\linewidth]{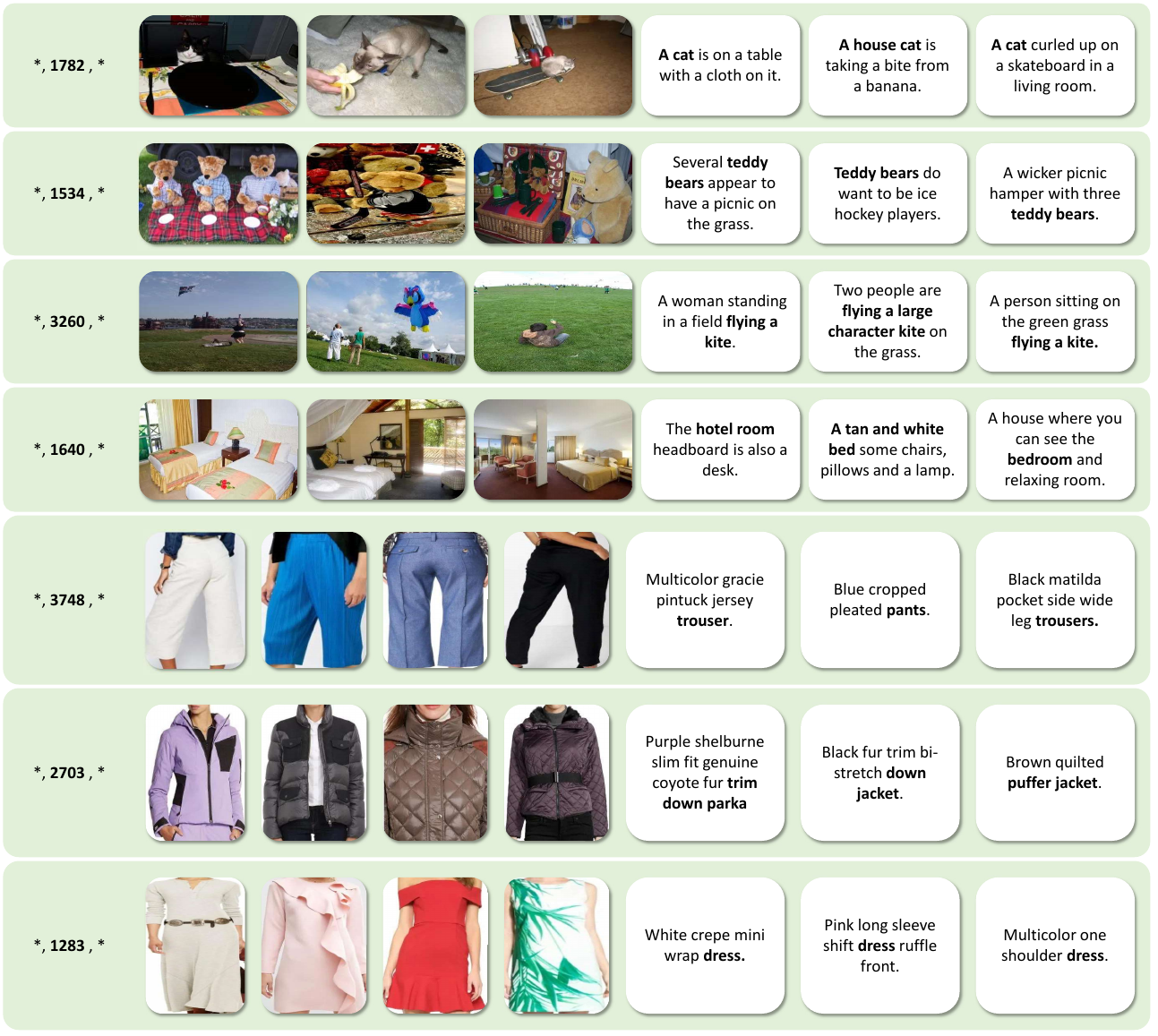}
\caption{Examples of second-level codes in the modality-decoupled semantic quantization. This level captures coarse semantics, such as primary objects or key scenes, with rows representing scenes from COCO and Fashion200K datasets.} 
\label{supfig:level2}
\end{figure*}

The \textbf{second level encodes} broad semantic concepts, capturing \emph{primary objects} or \emph{key scenes} shared across multimodal data. As shown in Fig.~\ref{supfig:level2}, examples include \texttt{1782} (\ie, a cat), grouping examples featuring cats in various contexts, such as lying on tables, eating bananas, or curling on skateboards. Other examples include \texttt{1534} (\ie, teddy bears), highlighting scenes like picnics or playful activities, and \texttt{3260} (\ie, flying a kite), which captures shared actions across different settings. Similarly, \texttt{1640} (\ie, hotel room) clusters scenes with shared elements like beds and lamps. These groupings extend naturally to other domains, categorizing items like dresses, trousers, and jackets based on shared object types.

\begin{figure*} [!t]
\centering
\includegraphics[width = 1.0\linewidth]{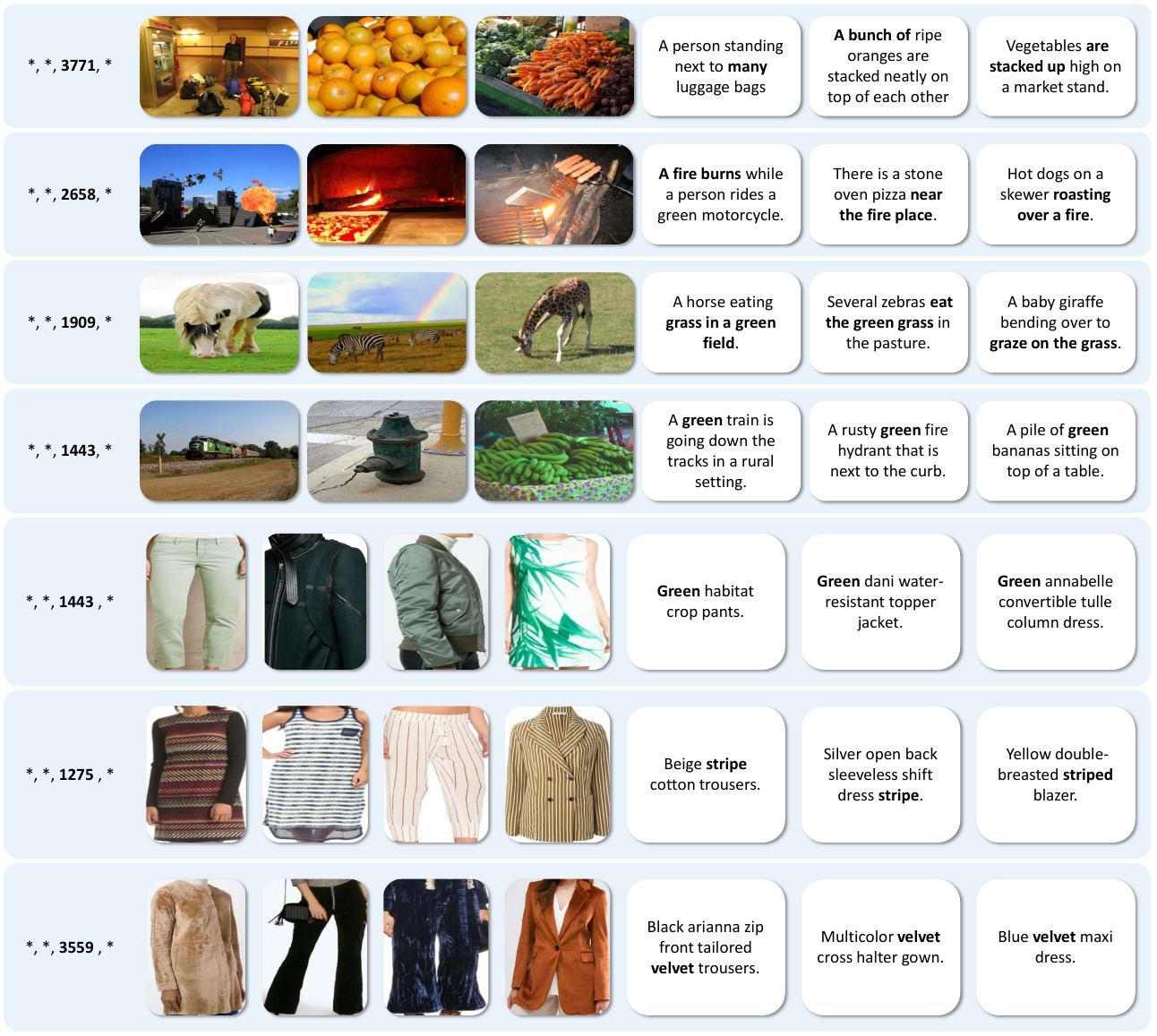}
\caption{Examples of third-level codes in the modality-decoupled semantic quantization. This level captures finer semantic attributes, such as object properties, material characteristics, or detailed patterns, across COCO and Fashion200K datasets.} 
\label{supfig:level3}
\end{figure*}

The \textbf{third-level codes} refine semantics by focusing on \emph{attributes} such as material, color, and patterns. Fig.~\ref{supfig:level3} illustrates these details. In COCO, \texttt{3771} (\ie, a bunch of) groups collections of items like stacked oranges, vegetables, or bananas, emphasizing grouping semantics. Similarly, \texttt{1443} (\ie, green) identifies objects prominently featuring green, such as train, fire hydrants, and bananas. In Fashion200K, \texttt{1443} (\ie, green) also highlights garments sharing the color green, while \texttt{1275} (\ie, striped clothing) focuses on items with striped patterns, such as blazers and trousers. Lastly, \texttt{3559} (\ie, velvet) captures items made of velvet material, regardless of the type of clothing, showcasing material-specific details.

\begin{figure*} [!t]
\vspace{-5mm}
\centering
\includegraphics[width = 1.0\linewidth]{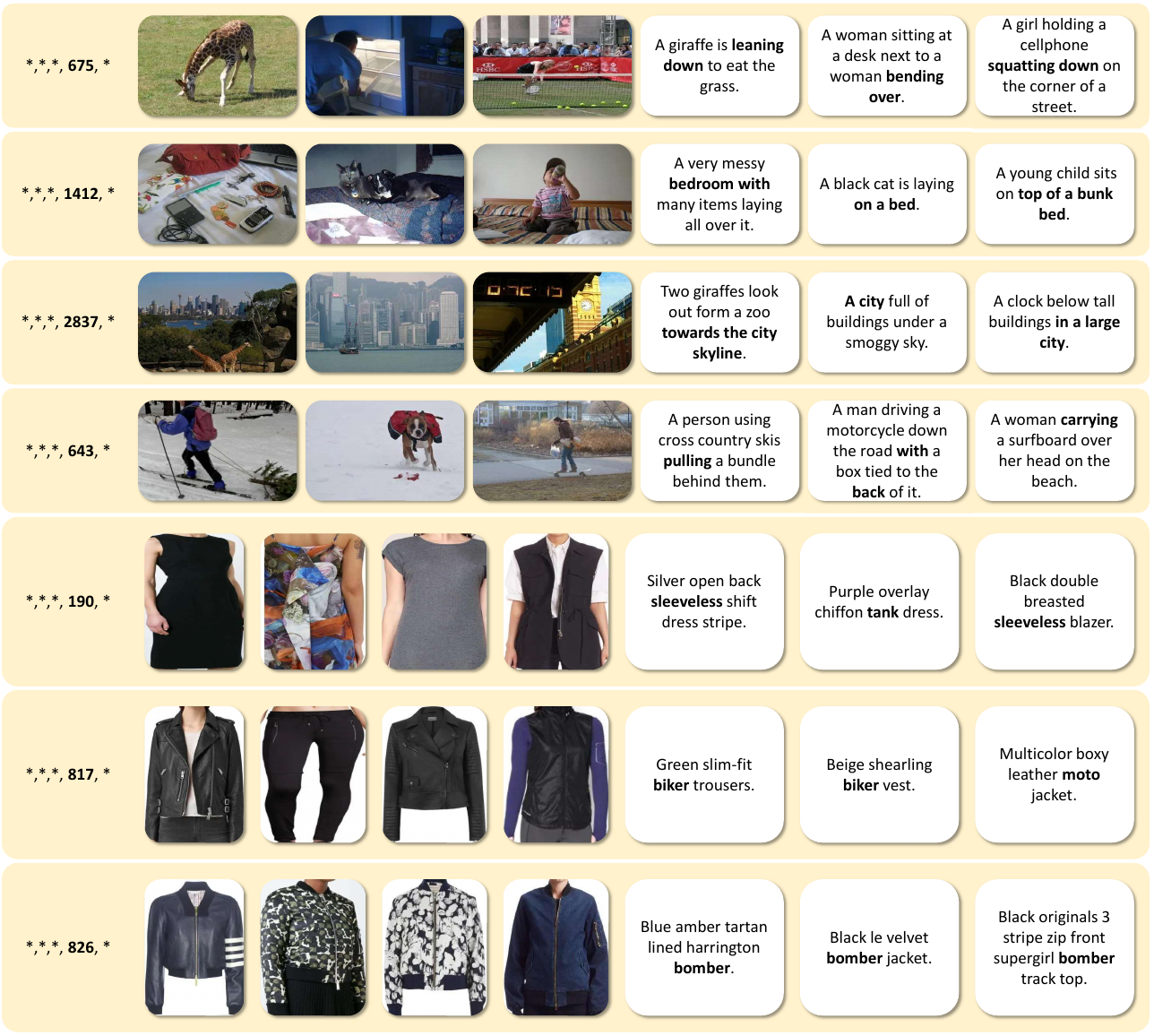}
\vspace{-2mm}
\caption{Examples of fourth-level codes in the modality-decoupled semantic quantization. This level captures highly fine-grained semantics, such as specific actions, positions, nuanced object details, or intricate clothing features.} 
\label{supfig:level4}
\vspace{-5mm}
\end{figure*}

The \textbf{fourth-level codes} capture \emph{highly fine-grained semantics}, such as specific actions, positions, and intricate design features. Fig.~\ref{supfig:level4} provides examples from COCO, including \texttt{675} (\ie, leaning down), which groups scenes featuring subjects leaning, such as giraffes eating grass or people bending over. Similarly, \texttt{1412} (\ie, in-bedroom) emphasizes indoor bedroom settings, capturing nuanced elements beyond generic room scenes. Furthermore, \texttt{643} (\ie, carrying) captures actions involving carrying objects, such as individuals carrying suitcases or animals transporting items. In Fashion200K, codes like \texttt{190} (\ie, sleeveless style), \texttt{817} (\ie, biker style), and \texttt{826} (\ie, bomber style) reflect fine-grained characteristics of garments, such as sleeveless cuts, biker styles, or specific jacket designs.

While the examples showcase the first four levels, the quantization process extends further to encode increasingly fine details, enriching semantic representation. Although these examples primarily showcase COCO and Fashion200K data, the quantization framework is designed to generalize across datasets. Shared semantics, such as \texttt{1443} (\ie, green) in second-level remain consistent across different domains, highlighting the universal nature of the code structure. This capability ensures consistent capturing and alignment of similar semantics, irrespective of the dataset.
These properties enable the decoder in our GENIUS framework to seamlessly map multimodal data to their corresponding codes. As a result, by leveraging this structured and interpretable quantization, GENIUS achieves not only high retrieval performance but also ensures generalization across a wide range of tasks, spanning various modalities and domains.

\end{document}